\documentclass[prd,twocolumn,superscriptaddress,altaffilletter,showpacs,nofootinbib]{revtex4}
\usepackage[dvips]{graphicx}
\usepackage{amsmath}

\newcommand{\be}{\begin{equation}}
\newcommand{\ee}{\end{equation}}
\newcommand{\bea}{\begin{eqnarray}}
\newcommand{\eea}{\end{eqnarray}}
\newcommand{\der}{\partial}
\newcommand{\vphi}{\varphi}

\begin{document}

\title{The Chameleon Effect in the Jordan Frame of the Brans--Dicke Theory}

\author{Israel Quiros}\email{iquiros@fisica.ugto.mx}\affiliation{Dpto. Ingenier\'ia Civil, Divisi\'on de Ingenier\'ia, Universidad de Guanajuato, Gto., M\'exico.}

\author{Ricardo Garc\'{\i}a-Salcedo}\email{rigarcias@ipn.mx}\affiliation{CICATA - Legaria del Instituto Polit\'ecnico Nacional, 11500, M\'exico, D.F., M\'exico.}

\author{Tame Gonzalez}\email{tamegc72@gmail.com}\affiliation{Dpto. Ingenier\'ia Civil, Divisi\'on de Ingenier\'ia, Universidad de Guanajuato, Gto., M\'exico.}

\author{F. Antonio Horta-Rangel}\email{anthort@hotmail.com}\affiliation{Dpto. Ingenier\'ia Civil, Divisi\'on de Ingenier\'ia, Universidad de Guanajuato, Gto., M\'exico.}

\date{\today}

\begin{abstract} In this paper we investigate the chameleon effect in the different conformal frames of the Brans--Dicke theory. Given that, in the standard literature on the subject, the chameleon is described in the Einstein frame almost exclusively, here we pay special attention to the description of this effect in the Jordan and in the string frames. It is shown that, in general, terrestrial and solar system bounds on the mass of the BD scalar field, and bounds of cosmological origin, are difficult to reconcile at once through a single chameleon potential. We point out that, in a cosmological context, provided that the effective chameleon potential has a minimum within a region of constant density of matter, the Brans--Dicke theory transmutes into general relativity with a cosmological constant, in that region. This result, however, can be only locally valid. In cosmological settings de Sitter--general relativity is a global attractor of the Brans--Dicke theory only for the quadratic potential $V(\phi)=M^2\phi^2$, or for potentials that asymptote to $M^2\phi^2$.\end{abstract}

\maketitle


\section{introduction}\label{intro-sec}

The Brans--Dicke (BD) theory of gravity \cite{bd} represents the simplest modification of general relativity (GR): in addition to the 10 degrees which are associated with the metric tensor $g_{\mu\nu}$, a new scalar degree of freedom is also responsible for propagating gravity. This theory has been cornerstone for a better understanding of several other modifications of general relativity, such as the $f(R)$ theories of gravity \cite{sotiriou-rev}. In contrast to Einstein's GR, the BD theory is not a fully geometrical theory of gravity, since, while one of the carriers of the gravitational field: the metric tensor, defines the metric properties of the spacetime, the scalar field $\phi$, which modifies the local strength of the gravitational interactions through the effective gravitational coupling $G_\text{eff}\propto\phi^{-1}$, is a non-geometric field. 

Although many aspects of BD theory have been well-explored in the past \cite{maeda-book, faraoni-book}, other aspects have been cleared up just recently. Thanks to the chameleon effect \cite{cham, cham-khoury, cham-tamaki, cham-mota, cham-olive, cosmo-cham, cham-5-force, cham-rad, fdr-cham, bd-cham, rev-cham}, for instance, it was just recently understood that the experimental bounds on the BD coupling parameter $\omega_\textsc{bd}$, which were set up through experiments in the solar system, might not apply in the large cosmological scales if consider BD theory with a potential. According to the chameleon effect, the effective mass of the scalar field $m_\phi$, depends on the background energy density of the environment: In the large cosmological scales where the background energy density is of the order of the critical density $\rho_\text{crit}\sim 10^{-31}$ g/cm$^3$, the effective mass is very small $m_\phi\sim H_0\sim 10^{-33}$ eV, so that the scalar field has impact in the cosmological dynamics. Meanwhile, in the solar system, where the averaged energy density of the environment is huge compared with $\rho_\text{crit}$, the effective mass is large $m_\phi>1$ mm$^{-1}$ ($m_\phi> 10^{-3}$ eV), so that the Yukawa--like contribution of the scalar field to the gravitational interaction $\propto e^{-m_\phi r}/r$, is short-ranged, leading to an effective screening of the scalar field in the solar system.  

There is one aspect of the chameleon effect that has not yet been discussed in detail. If take a look at the existing bibliography on this subject one immediately finds that this effect is almost exclusively described in the Einstein frame (EF), where the scalar field (the chameleon) is minimally coupled to the curvature scalar, at the cost of being non--minimally coupled to the matter sector of the action. A detailed description of this effect either in the Jordan frame (JF) or in the string frame (SF) is yet lacking. Since the chameleon effect is apparent in the density dependence of the dilaton's mass, we think that the absence of appropriate discussion in the JF/SF, is due to the unconventional way which the self-interaction potential of the dilaton arises in the JF/SF Klein--Gordon (KG) equation, that governs its dinamics. 

In the present paper we shall try to fill the gap and the focus will be in the description of the chameleon effect in the Jordan and in the string frames of the BD theory with a potential (see sections \ref{setup-sec}, \ref{bd-mass-sec}, \ref{cham-mass-sec}, and \ref{dilaton-mass-sec}). Although the very warped subject of the conformal transformations controversy will not be discussed here, the consequences for the chameleon effect of the most widespread viewpoints on the (in)equivalence of the JF/SF and the EF, will be briefly discussed (see sections \ref{frames-sec}, \ref{ef-sec}, and \ref{conf-frames-sec}). 

In section \ref{gr-attractor-sec}, we will show that, in a cosmological context, provided that the effective chameleon potential has a minimum within a region of constant matter density, the GR--de Sitter solution can be, at most, either a local attractor or a saddle point of the Brans--Dicke theory within that region. In contrast, as it has been shown in \cite{hrycyna, bd-quiros} by means of the tools of the dynamical systems theory, in a cosmological setting the GR--de Sitter solution can be a global attractor of the BD theory exclusively for the quadratic potential: $V(\phi)=M^2\phi^2$, or for any BD potential that asymptotes to the quadratic one, $V(\phi)\rightarrow M^2\phi^2$ (see the section \ref{dyn-syst-sec}).\footnote{There are found several works on the de Sitter (inflationary) solutions within the scalar--tensor theory in the bibliography, in particular within the Brans--dicke theory, but just for illustration here we mention those in the references \cite{barrow-ref, odintsov-ref}.} 

In section \ref{estimate-sec} of this paper (see also the section \ref{ccp-sec}), we will show that the expectation that the chameleon effect might have had for the Brans--Dicke theory, specially for the relaxation of the stringent lower bounds on the value of the BD coupling constant $\omega_\textsc{bd}>40000$, is unjustified in general. It will be shown in sections \ref{eje-sec} and \ref{dilaton-mass-sec} (see the discussion in \ref{discu-conclu-sec}), that several BD models of importance for cosmology: those whose dynamics is driven by the quadratic potential $V(\phi)\propto\phi^2$, or by potentials that asymptote to the quadratic one,\footnote{As shown in section \ref{dyn-syst-sec}, these are the only BD cosmological models that have the $\Lambda$CDM model as a global attractor.} do not develop the chameleon effect. This latter effect is primordial for the relaxation of the bounds on $\omega_\textsc{bd}$, since, what one expects is that, thanks to the chameleon effect, the cosmological BD field is screened from terrestrial and solar system experimentation, even for $\omega_\textsc{bd}\sim 1$.


\section{basic setup}\label{setup-sec}

Here we assume the Brans--Dicke theory \cite{bd} with the potential, to dictate the dynamics of gravity and matter. In the Jordan frame it is depicted by the following action:

\bea S^\phi_\textsc{jf}=\int d^4x\sqrt{|g|}\left\{\phi R-\frac{\omega_\textsc{bd}}{\phi}(\der\phi)^2-2V+2{\cal L}_m\right\},\label{bd-action}\eea where $(\der\phi)^2\equiv g^{\mu\nu}\der_\mu\phi\der_\nu\phi$, $V=V(\phi)$ is the scalar field self-interaction potential, $\omega_\textsc{bd}$ is the BD coupling parameter, and ${\cal L}_m={\cal L}_m(\chi,\der\chi,g_{\mu\nu})$ is the Lagrangian density of the matter degrees of freedom, collectively denoted by $\chi$. Unless the contrary is specified, the natural units $8\pi G=1/M_\textsc{PL}^2=c=1$, are adopted. The field equations which are derived from (\ref{bd-action}) are the following:

\bea &&G_{\mu\nu}=\frac{\omega_\textsc{bd}}{\phi^2}\left[\der_\mu\phi\der_\nu\phi-\frac{1}{2}g_{\mu\nu}\left(\der\phi\right)^2\right]-g_{\mu\nu}\frac{V}{\phi}\nonumber\\
&&\;\;\;\;\;\;\;\;\;\;\;\;\;\;+\frac{1}{\phi}\left(\nabla_\mu\der_\nu\phi-g_{\mu\nu}\nabla^2\phi\right)+\frac{1}{\phi}\,T^{(m)}_{\mu\nu},\label{bd-feq}\\
&&\nabla^2\phi=\frac{2}{3+2\omega_\textsc{bd}}\left(\phi\der_\phi V-2V+\frac{1}{2}\,T^{(m)}\right),\label{bd-kg}\eea where $G_{\mu\nu}=R_{\mu\nu}-g_{\mu\nu}R/2$, is the the Einstein's tensor, $\nabla^2=g^{\mu\nu}\nabla_\mu\nabla_\nu$, is the D'Alembertian operator, and $$T^{(m)}_{\mu\nu}=-\frac{2}{\sqrt{|g|}}\frac{\der\left(\sqrt{|g|}\,{\cal L}_{m}\right)}{\der g^{\mu\nu}},$$ is the conserved stress-energy tensor of the matter degrees of freedom $$\nabla^\mu T^{(m)}_{\mu\nu}=0.$$

It is also convenient to rescale the BD scalar field and, consequently, the self-interaction potential:

\bea \phi=e^\vphi,\;V(\phi)=e^\vphi\,U(\vphi),\label{vphi}\eea so that, the action (\ref{bd-action}) is transformed into the string frame BD action:

\bea S_\textsc{sf}^\vphi=\int d^4x\sqrt{|g|}e^\vphi\left\{R-\omega_\textsc{bd}(\der\vphi)^2-2U+2e^{-\vphi}{\cal L}_m\right\}.\label{dilaton-action}\eea The following motion equations are obtained from (\ref{dilaton-action}):

\bea &&G_{\mu\nu}=\left(\omega_\textsc{bd}+1\right)\left[\der_\mu\vphi\der_\nu\vphi-\frac{1}{2}g_{\mu\nu}(\der\vphi)^2\right]\nonumber\\
&&\;\;\;\;\;\;\;\;\;\;\;\;\;\;\;\;\;\;\;\;\;\;\;\;\;\;\;-g_{\mu\nu}\left[\frac{1}{2}(\der\vphi)^2+U(\vphi)\right]\nonumber\\
&&\;\;\;\;\;\;\;\;\;\;\;\;\;\;\;\;\;\;\;\;+\nabla_\mu\der_\nu\vphi-g_{\mu\nu}\nabla^2\vphi+e^{-\vphi}T^{(m)}_{\mu\nu},\label{dilaton-feq}\\
&&\nabla^2\vphi+(\der\vphi)^2=\frac{2\left[\der_\vphi U-U+\frac{e^{-\vphi}}{2}T^{(m)}\right]}{3+2\omega_\textsc{bd}},\label{dilaton-kg}\eea where $\nabla^2\equiv g^{\mu\nu}\nabla_\mu\der_\nu$, $G_{\mu\nu}=R_{\mu\nu}-g_{\mu\nu}R/2$, and, as before, $T^{(m)}_{\mu\nu}$ is the (conserved) stress-energy tensor of the matter degrees of freedom: $\nabla^\mu T^{(m)}_{\mu\nu}=0$. 

It is clear that the equations (\ref{bd-action}), (\ref{bd-feq}), (\ref{bd-kg}), in the one hand, and (\ref{dilaton-action}), (\ref{dilaton-feq}), (\ref{dilaton-kg}), in the other, transform into each other under the redefinitions in Eq. (\ref{vphi}).


\section{the klein-gordon equation and the mass of the scalar field}\label{bd-mass-sec}

The mass (squared) of the BD scalar field can be computed with the help of the following equation \cite{mass}: 

\bea m_\phi^2=\frac{2}{3+2\omega_\textsc{bd}}\left[\phi \der^2_\phi V(\phi)-\der_\phi V(\phi)\right].\label{bd-mass}\eea This mass is the one which is associated with a Yukawa--like term $\phi(r)\propto\exp(-m_\phi r)/r$, when the Klein--Gordon equation (\ref{bd-kg}) is considered in the weak--field, slow--motion regime, and in the spherically symmetric case. The way in which this result is obtained can be found in the reference \cite{mass} in all details. However, for completeness of our exposition, here we shall explain the main reasoning line behind this result.

In general for a scalar field which satisfies the standard KG equation 

\bea \nabla^2\phi=\der_\phi V_\text{eff}+S,\label{kg-eq}\eea where $V_\text{eff}=V_\text{eff}(\phi)$ is the effective self-interaction potential of the scalar field $\phi$, and $S$ is a source term ($S$ does not depend on $\phi$), the effective mass squared of the scalar field is defined by $m^2_\phi=\der^2_\phi V_\text{eff}$. The problem with this definition is that the BD scalar field does not satisfy the usual KG equation, but the one in Eq. (\ref{bd-kg}), where the self-interaction potential of the BD field appears in an unconventional way. In order to fix this problem, one notices that, if introduce the effective potential \cite{mass}

\bea V_\text{eff}(\phi)=\frac{2}{3+2\omega_\textsc{bd}}\left[\phi V(\phi)-3\int d\phi V(\phi)\right],\label{eff-pot}\eea so that $$\der_\phi V_\text{eff}=\frac{2}{3+2\omega_\textsc{bd}}\left[\phi\der_\phi V(\phi)-2V(\phi)\right],$$ then, the KG equation (\ref{bd-kg}) can be rewritten in the more conventional way: 

\bea \nabla^2\phi=\der_\phi V_\text{eff}+\frac{1}{3+2\omega_\textsc{bd}}\,T^{(m)},\label{kg-source}\eea where the second term in the right-hand side (RHS) of this equation, is the source term which does not depend explicitly on the $\phi$--field.

\subsection{Perturbations around the minimum of the effective potential}\label{perts-subsec} 

Usually the details of the derivation that leads to the appearance of an effective mass of the scalar field are not given. Here we want to include these details, since we want to make clear that, what one usually calls as an effective mass, is a concept that is linked with the oscillations of the field around the minimum of the effective potential, which propagate in spacetime. These oscillations, or excitations, are the ones that carry energy--momentum and, if required, may be quantized.

For simplicity consider the vacuum case $T^{(m)}_{\mu\nu}=0$ of Eq. (\ref{kg-source}). Let us assume next that $V_\text{eff}$ is a minimum at some $\phi_*$. Given that the equation (\ref{kg-source}) is non--linear, one may consider small deviations around $\phi_*$: $\phi=\phi_*+\delta\phi$ ($\delta\phi\ll 1$), then, up to terms linear in the deviation, one gets: $$\der_\phi V_\text{eff}(\phi)\approx\der_\phi V_\text{eff}(\phi_*)+\der^2_\phi V_\text{eff}(\phi_*)\,\delta\phi+...=m^2_*\delta\phi,$$ where $m^2_*\equiv \der^2_\phi V_\text{eff}(\phi_*)$, is the effective mass of the scalar field perturbations. 

Working in a flat Minkowski background (in spherical coordinates) $$ds^2=-dt^2+dr^2+r^2\left(d\theta^2+\sin^2\theta d\phi^2\right),$$ which amounts to ignoring the curvature effects and the backreaction of the scalar field perturbations on the metric, and imposing separation of variables, the perturbations $$\delta\vphi=\delta\vphi(t,r)=\sum_n e^{-i\omega_n t}\psi_n(r),$$ where $\omega_n$ is the angular frequency of the oscillations of the $n$-th excitation of the field, obey the Helmholtz equation: 

\bea \frac{d^2\psi_n}{dr^2}+\frac{2}{r}\frac{d\psi_n}{dr}+{\bf k}_n^2\psi_n=0,\label{helmholtz-eq}\eea where ${\bf k}_n^2=\omega^2_n-m^2_*$, is the wave--number squared. Eq. (\ref{helmholtz-eq}) is solved by the spherical waves: $$\psi_n(r)=C_n\frac{e^{i|{\bf k}_n|r}}{r},$$ where the $C_n$--s are integration constants. For $|\omega_n|<m_*$, the Eq. (\ref{helmholtz-eq}) has the Yukawa--type solution:

\bea \psi_n(r)=C_n\frac{e^{-\sqrt{m^2_*-\omega_n^2}\,r}}{r}.\label{yukawa-sol}\eea 

It is understood that the modes with the lowest energies $E_n=\hbar\,\omega_n=\omega_n\ll m_*$, are the ones which are more easily excited in the small oscillations approximation, and, hence, are the prevailing ones. In particular, $\omega_0\ll m_*$, so that

\bea \psi(r)\sim\psi_0(r)=C\frac{e^{-m_*\,r}}{r}.\label{0-mode}\eea These lowest order modes are the ones with the shortest effective Compton wave length $\lambda_*\approx m^{-1}_*$, and are the ones which decide the range of the Yukawa-type interaction, i. e., these are the ones which decide the effective screening of the $\phi$--field.

In what follows, as it is usually done in the related bibliography, when we talk about perturbations of the scalar field around the minimum of the effective potential, we ignore the oscillations in time by assuming the static situation, etc. This amounts, effectively, to ignoring all of the higher--energy excitations of the field. This assumption bears no consequences for the qualitative analysis. However, once a friction term $\propto\dot\phi$ arises (the dot accounts for the time derivative), for instance, in a cosmological context, the oscillations of the field in time, around the minimum, are necessarily to be considered (see section \ref{gr-attractor-sec}).

\subsection{The field--theoretical mass}

The above analysis suggests that one may introduce a general definition of the mass (squared) of the BD scalar field $m^2_\phi\rightarrow m^2_*=\der^2_\phi V_\text{eff}(\phi_*)$, $$m^2_\phi=\der^2_\phi V_\text{eff}(\phi)=\frac{2}{3+2\omega_\textsc{bd}}\left[\phi \der^2_\phi V(\phi)-\der_\phi V(\phi)\right],$$ which is just Eq. (\ref{bd-mass}) at the beginning of this section. 

What if the effective potential $V_\text{eff}$ has no minimums at all? The quartic potential $V(\phi)=\lambda\phi^4$, for instance, has a minimum at $\phi=0$, however, the corresponding effective potential: 

\bea V_\text{eff}(\phi)=\frac{4\lambda\,\phi^5}{5(3+2\omega_\textsc{bd})},\label{eff-q-pot}\eea has no minimums. In this case, our understanding of what an effective mass means, might have no meaning at all. In particular, the screened Coulomb--type potential (the mentioned Yukawa-like solution), being the most relevant physical manifestation of a massive propagator, might not arise. The corresponding ``mass'' in Eq. (\ref{bd-mass}) would be just a useful field theoretical construction with the dimensions of mass, no more.

In spite of this, following the most widespread point of view on the correct definition of the mass of the BD field in the Jordan frame \cite{mass}, here we shall consider that, in general, i. e., even away from the minimum of the effective potential, the field parameter $m^2_\phi$ given by Eq. (\ref{bd-mass}), represents the effective mass (squared) of the scalar field. In this paper, in order to differentiate the mentioned field theoretic parameter from an actual effective mass (the one with consequences for fifth--force experiments), we shall call the latter as ``effective mass'', while the former as ``effective field--theoretical mass'' (see the related discussion in section \ref{eff-ft-mass-subsec}).


\section{the chameleon mass}\label{cham-mass-sec}

Due to the chameleon mechanism, the screening effect may arise even if the effective potential $V_\text{eff}$ does not develop minimums. As we shall see, all what one needs is to include the source term in the RHS of the KG equation (\ref{kg-eq}), within a redefined effective potential, which we shall call effective chameleonic potential: 

\bea V_\text{ch}=V_\text{eff}+\phi S\;\Rightarrow\;\nabla^2\phi=\der_\phi V_\text{ch}=\der_\phi V_\text{eff}+S.\label{cham-pot-idea}\eea 

Usually the chameleon effect is discussed, exclusively, in the Einstein frame formulation of the Brans--Dicke theory, where the scalar field couples directly with the matter degrees of freedom \cite{cham, cham-khoury, cham-tamaki, cham-mota, cham-olive, cosmo-cham, cham-5-force, cham-rad, fdr-cham, bd-cham, rev-cham}. Due to the non--trivial way which the self--interaction potential enters in the KG equation, the discussion of the chameleon effect in the Jordan frame formulation of BD theory seems more obscure than in the Einstein frame.

Here we shall show that, regardless of the unconventional form of the potential in the BD--KG equation (\ref{bd-kg}), the chameleon effect can be discussed in the Jordan frame as well, if in Eq. (\ref{bd-kg}) introduce the following definition of the effective chameleon potential:

\bea &&V_\text{ch}(\phi)=V_\text{eff}(\phi)+\frac{1}{3+2\omega_\textsc{bd}}\,\phi\,T^{(m)}=\nonumber\\
&&\frac{2}{3+2\omega_\textsc{bd}}\left[\phi V(\phi)-3\int d\phi V(\phi)+\frac{1}{2}\,\phi\,T^{(m)}\right],\label{bd-cham-pot}\eea so that 

\bea \der_\phi V_\text{ch}=\frac{2}{3+2\omega_\textsc{bd}}\left(\phi\der_\phi V-2V+\frac{1}{2}\,T^{(m)}\right),\label{der-bd-cham--pot}\eea coincides with the RHS of the KG Eq. (\ref{bd-kg}), and the latter can be written in the form of the conventional KG equation without a source: $\nabla^2\phi=\der_\phi V_\text{ch}.$ 

As in the standard case, the effective mass squared: $m^2_{\phi_*}=\der^2_\phi V_\text{ch}(\phi_*),$ may be defined for the small perturbations of the BD scalar field around the minimum $\phi_*$ of the chameleon potential $V_\text{ch}$. Actually, under the assumption of spherical symmetry, given that $V_\text{ch}$ is a minimum at some $\phi_*$, if follow the procedure explained in the former section \ref{bd-mass-sec}, in the weak--field and low--velocity limit (basically the case when the curvature effects and the backreaction on the metric are ignored): $$\frac{d^2(\delta\phi)}{dr^2}+\frac{2}{r}\frac{d(\delta\phi)}{dr}=m^2_{\phi_*}\delta\phi,$$ where $$m^2_{\phi_*}=\der^2_\phi V_\text{ch}(\phi_*),$$ is the effective mass of the perturbations around the minimum of the chameleon potential. 

We recall that, as shown in the section \ref{perts-subsec}, although when dealing with the chameleon effect we care only about the spatial deviation about the minimum $\delta\phi(r)$, as a means to linearize the Klein--Gordon equation, in general these deviations are also time--dependent, so that what we have are time--dependent perturbations around the minimum of the chameleon potential. These may be viewed as periodic oscillations of the BD field about the minimum, and the resulting effective mass can be interpreted as the mass of the corresponding scalar excitations propagating in a flat background.

Solving for the above Helmholtz equation, one has for $\phi(r)=\phi_*+\delta\phi(r)$, the following solution:$$\phi(r)=\phi_*+C_1\frac{e^{-m_{\phi_*} r}}{r}+C_2\frac{e^{m_{\phi_*} r}}{r},$$ where $C_1$ and $C_2$ are integration constants which we can determine through the boundary conditions. If assume, for instance, that $\phi(r)$ tends asymptotically to a constant value $\phi_\infty$: $$\lim_{r\rightarrow\infty}\phi(r)=\phi_\infty\;\Rightarrow\;\phi(r)=\phi_\infty+C_1\frac{e^{-m_{\phi_*} r}}{r}.$$ 

Depending on the physical situation at hand, other boundary conditions are required in order to fix the remaining constants $C_1$ and $C_2$. For instance, if assume regularity of the solution at the origin $r=0$, then $C_1=-C_2$, $$\phi(r)=\phi_*+C_2\frac{\sinh(m_{\phi_*} r)}{r},$$ so that $\phi(0)=\phi_*+C_2 m_*$, etc. 

As we shall see, the interesting thing here is that the effective chameleon mass $m_{\phi_*}=m_{\phi_*}(\rho)$, is a function of the surrounding density $\rho$. This property of the effective mass of the BD scalar field perturbations is what is called, primarily, as the chameleon effect.

\subsection{On the matter density} 

An interesting aspect of the BD chameleon effect is related with the fact that the density of matter $\rho$ in the argument of the effective mass: $m_{\phi_*}(\rho)$, is the density measured by co--moving observers (with four--velocity $\delta^\mu_0$) in the JF--BD theory: $\rho=T^{(m)}_{\mu\nu}\delta^\mu_0\delta^\nu_0$. This is, besides, the density of matter that is conserved in the JF (also in the SF) formulation of the Brans--Dicke theory. 

This is to be contrasted with the original chameleon effect of Ref. \cite{cham}, where the physically meaningful matter density $\rho_i$ is not the one measured by co--moving observers in the EF (this is not the conserved one in this frame), neither in the conformal one, but a density which does not depend on the dilaton. Actually, in \cite{cham} the matter fields couple to the conformal metric $g^{(i)}_{\mu\nu}=\exp(2\beta_i\phi/M_\textsc{pl})\,g_{\mu\nu}$, while the density of the non--relativistic fluid measured by EF co--moving observers is denoted by $\tilde\rho_i$. It is assumed that what matters is the $\phi$--independent density $\rho_i=\tilde\rho_i\exp(3\beta_i\phi/M_\textsc{pl})$, which is the one conserved in the EF. While this choice may not be unique, in the Jordan frame of the Brans--Dicke theory (the same for the SF), one does not have this ambiguity: the matter density measured by JF(SF) co--moving observers $\rho$, is the one conserved in the Jordan/string frames and, additionally, does not depend on the BD--field.


\section{the bd chameleon: examples}\label{eje-sec}

In this section we shall to illustrate how the chameleon effect arises in the Jordan frame of the Brans--Dicke theory, by exploring a pair of examples.

\subsection{The quartic potential as an example}\label{quartic-sub-sec}

In order to illustrate our reasoning, let us choose the example with the quartic potential \cite{cham-khoury}: 

\bea V(\phi)=\lambda\phi^4,\label{q-pot}\eea where we assume that the free parameter $\lambda\geq 0$, is a non-negative constant. In this case, as said, the effective potential (\ref{eff-q-pot}): $V_\text{eff}(\phi)\propto\phi^5$, does not develop minimums. Yet the corresponding chameleon potential (\ref{bd-cham-pot}):

\bea &&V_\text{ch}(\phi)=\frac{4}{3+2\omega_\textsc{bd}}\left[\frac{\lambda}{5}\,\phi^5+\frac{T^{(m)}}{4}\,\phi\right]\nonumber\\
&&\;\;\;\;\;\;\;\;\;\;\;\;\;\;\;\;\;=\frac{4}{3+2\omega_\textsc{bd}}\left[\frac{\lambda}{5}\,\phi^5-\frac{\rho}{4}\,\phi\right],\label{bd-cham-pot-q}\eea where we have assumed a homogeneous, pressureless dust background: $T^{(m)}=-\rho$, can have a minimum if $\omega_\textsc{bd}\geq-3/2$. Actually, at the value $$\phi_*=\left(\frac{\rho}{4\lambda}\right)^{1/4},$$ the derivatives of the above chameleon potential $$\der_\phi V_\text{ch}(\phi_*)=0,\;\der^2_\phi V_\text{ch}(\phi_*)=\frac{16\lambda}{3+2\omega_\textsc{bd}}\left(\frac{\rho}{4\lambda}\right)^{3/4}.$$ Hence, provided that $\omega_\textsc{bd}\geq-3/2$, since $\der^2_\phi V_\text{ch}(\phi_*)>0$, the chameleon potential $V_\text{ch}$ in Eq. (\ref{bd-cham-pot-q}), is a minimum at $\phi_*$. In this case we can identify a physically meaningful effective mass of the BD field:

\bea m^2_{\phi_*}=\der^2_\phi V_\text{ch}(\phi_*)=\frac{16\lambda}{3+2\omega_\textsc{bd}}\left(\frac{\rho}{4\lambda}\right)^{3/4}.\label{phys-bd-mass}\eea 

We want to underline that, in general, $\rho$ can be a function of the spacetime point $\rho=\rho(x)$, however, in most applications the function $\rho(x)$ is assumed piece--wise constant. For instance, one may imagine an spherical spatial region of radius $R$, filled with a static fluid with homogeneous and isotropic constant density $\rho_0$, and surrounded by a fluid with a different (also homogeneous) constant density $\rho_\infty$, so that: \[\rho(r)=\left\{\begin{array}{ll} \rho_0 & \mbox{if $r\leq R$};\\ \rho_\infty & \mbox{if $r\gg R$}.\end{array} \right.\] 

In such a case the effective mass $m_{\phi_*}$ of the BD scalar field would be one for modes propagating inside the spherical region $m_{\phi_0}$, and another different value $m_{\phi_\infty}$, for scalar modes propagating outside of (far from) the spherical region: $$m_{\phi_0}=\sqrt{\der^2_\phi V_\text{ch}(\phi_0)},\;m_{\phi_\infty}=\sqrt{\der^2_\phi V_\text{ch}(\phi_\infty)}.$$ For the quartic potential, in particular, one would have that: $$m_{\phi_0}=\frac{2(4\lambda)^{1/8}\,\rho^{3/8}_0}{\sqrt{3+2\omega_\textsc{bd}}},\;m_{\phi_\infty}=\frac{2(4\lambda)^{1/8}\,\rho^{3/8}_\infty}{\sqrt{3+2\omega_\textsc{bd}}},$$ inside and outside of the spherical region of radius $R$, respectively.

In case the the gravitational configuration of matter were given by a point--dependent density profile $\rho=\rho(x)$, such as, for instance, in a cosmological context where $\rho=\rho(t)$ ($t$ is the cosmic time), the effective chameleon mass $m_{\phi_*}$ were point-dependent as well. However, as it is well known, the masses of point particles in the JF/SF formulations of the BD theory, are constants by definition. Otherwise, these particles would not follow geodesics of the JF/SF metric. In general, coexistence of particles of constant mass and particle excitations with point--dependent mass, bring about problems with the equivalence principle. Besides, 
if $\rho=\rho(x)$, then, the resulting effective mass will be a field--theoretical construction which, as it will be shown in section \ref{eff-ft-mass-subsec}, has an anomalous behavior under the conformal transformation of the metric.

In order to evade any possible discussion on the equivalence principle, or on the anomalous behavior of the effective field--theoretical chameleon mass under the conformal transformations of the metric, here we adopt the most widespread handling of the chameleon effect, and we assume that the density of matter has piece--wise constant profile in the sense explained above. In section \ref{gr-attractor-sec} we will discuss more on this delicate subject.

\subsection{The quadratic monomial: the massless BD field}\label{q-pot-subsec}

One peculiar note about the effective chameleon potentials: if look at equation (\ref{bd-cham-pot}), one sees that the quadratic potential $V(\phi)=M^2\phi^2$, plays a singular role. Actually, if substitute this potential into (\ref{bd-cham-pot}), one obtains that the resulting chameleon potential $$V_\text{ch}(\phi)=\frac{\phi\,T^{(m)}}{3+2\omega_\textsc{bd}},$$ does not have a minimum. Besides, the second derivative vanishes: $\der^2_\phi V_\text{ch}=0$. This means that the quadratic potential does not generate the chameleon effect. The same is true for any potential that asymptotes to $\phi^2$, for instance $V(\phi)\propto\cosh(\lambda\phi)-1$. Even the effective field--theoretical mass squared (\ref{bd-mass}), vanishes for the quadratic monomial.

As a consequence, assuming the 'mexican hat' potential $$V(\phi)=-M^2\phi^2+\lambda\phi^4,$$ the corresponding effective and chameleon potentials (\ref{eff-pot}), and (\ref{bd-cham-pot}), will be given by: $$V_\text{eff}=\frac{4\lambda\,\phi^5}{5(3+2\omega_\textsc{bd})},\;V_\text{ch}=\frac{4}{3+2\omega_\textsc{bd}}\left[\frac{\lambda}{5}\,\phi^5-\frac{\rho}{4}\,\phi\right],$$ respectively, which coincide with the effective and chameleon potentials for the quartic monomial (\ref{q-pot}). This means that the quadratic monomial does not contribute neither to the effective nor to the chameleon potentials. 

In the case of a standard scalar field $\sigma$ whose dynamics is governed by the usual Kelin--Gordon equation $\nabla^2\sigma=\der_\sigma V(\sigma)$, the quadractic monomial $V(\sigma)\propto\sigma^2$, is also a singular potential in the sense that, it is the only potential for which the KG equation is a linear differential equation, i. e., the superposition principle is satisfied.


\section{the quartic potential: estimates}\label{estimate-sec}

Notice that, similar to the chameleon effect arising in the Einstein frame of the BD theory \cite{cham}, the mass squared of the BD field given by Eq. (\ref{phys-bd-mass}), i. e., the Jordan frame mass -- the one that determines the range of the Yukawa-like correction \cite{sotiriou} -- depends on the background energy density $m_{\phi_*}\propto\rho^{3/8}$. As it can be seen, this dependence of the mass of the scalar field on the ambient energy density improves the one in Ref. \cite{cham-khoury}: $m_\phi\propto\rho^{1/3}$, just by a fraction. 

In order to make estimates, let us write $$m_{\phi_*}=\frac{4^{5/8}\lambda^{1/8}}{\sqrt{3+2\omega_\textsc{bd}}}\left(\frac{\rho}{M^4_\textsc{pl}}\right)^{3/8}M_\textsc{pl},$$ or in ``user-friendly'' units (using the terminology of \cite{cham-khoury}): 

\bea m_{\phi_*}[\text{mm}^{-1}]\approx\frac{10\lambda^{1/8}}{\sqrt{3+2\omega_\textsc{bd}}}\left(\rho[\text{g/cm}^3]\right)^{3/8}.\label{estimate-eq}\eea 

Let us assume that the scalar field is immersed in the earth atmosphere with mean density $\rho^\text{atm}\approx 10^{-3}$ g/cm$^3$, then, provided that the millimeter range screening \cite{cham-khoury}: $(m_{\phi_*}^\text{atm})^{-1}\sim 1$ mm, is undertaken, from Eq. (\ref{estimate-eq}) it follows that $$\omega_\textsc{bd}\approx 1.6\,\lambda^{1/4}-1.5,$$ so that, if consider, for instance, that $\lambda\sim 1$, one gets that $\omega_\textsc{bd}\approx 0.1$ can be of order unity or smaller. What this means is that the BD theory may describe the gravitational phenomena with a coupling constant of order unity and, yet, the chameleon potential (\ref{bd-cham-pot-q}) may effectively screen the BD field from experiments that look for violation of the Newton's law, for distances above the millimeter.

The next question is whether the above potential can be a good candidate for cosmology as well. The ratio of the mass of the BD field measured at large cosmological scales, to the scalar field mass estimated in earth's atmosphere: 

\bea \frac{m_{\phi_*}^\text{cosm}}{m_{\phi_*}^\text{atm}}=\left(\frac{\rho^\text{crit}}{\rho^\text{atm}}\right)^{3/8}\approx 3\times 10^{-11},\label{estimate}\eea where we have taken into account that the critical energy density of the universe $\rho^\text{crit}\sim 10^{-31}$ g/cm$^3$. If consider the millimeter--range screening above, $(m^\text{atm}_{\phi_*})^{-1}\approx 1$mm $\Rightarrow\;m^\text{atm}_{\phi_*}\approx 10^{-4}$ eV, then the estimated mass of the cosmological BD scalar field $$m_{\phi_*}^\text{cosm}\approx 3\times 10^{-11} m_{\phi_*}^\text{atm}\approx 3\times 10^{-15}\;\text{eV},$$ is by some 18 orders of magnitude heavier than the expected value $m_\phi^\text{cosm}\sim H_0\sim 10^{-33}$ eV. Hence, if assume that the BD scalar field with a fixed potential $V(\phi)=\lambda\phi^4$, is effectively screened from solar system experimentation, the BD field would not have cosmological implications.

The ``reconciliation'' between terrestrial and cosmological bounds, at once, can be achieved by power--law potentials leading to BD chameleon mass, $m_{\phi_*}\propto(\rho)^{k/2}$, with the power $k\approx 29/14\approx 2.071$, or higher. Of course, the reconciliation is natural if, for instance, $m_{\phi_*}\propto\exp\rho$.

\subsection{Thin--shell effect}\label{thin-shell-subsec}

Our estimates above are unsatisfactory in many aspects. First of all, a lot of simplification has been made for sake of transparency of our analysis. For instance, the well known thin-shell effect \cite{cham}, which arises due to the non-linearity of the BD scalar field, and which is significant for large bodies, has not been considered in our analysis.\footnote{For a detailed exposition of the thin--shell effect we recommend \cite{cham-khoury} (See also the appendix \ref{appendix}).} Nevertheless, even if take into account the thin--shell effect, the physical implications of the huge difference between the cosmic and terrestrial mass scales: $m_{\phi_*}^\text{cosm}/m_{\phi_*}^\text{atm}\sim 10^{-11}$, can not be erased by the (thin--shell mediated) weakening of the effective coupling of the BD field to the surrounding matter. Actually, the additional contribution of the chameleon BD field to the Newtonian gravitational potential energy of a given mass $M_b$, is expressed by $$\Delta U^*_N\propto-\beta^{*2}_\text{eff}M_b\frac{e^{-r/\lambda_\text{eff}}}{r},$$ where $\beta^*_\text{eff}$ is the effective coupling of the chameleon field to the surrounding matter, and $\lambda_\text{eff}=m_{\phi_*}^{-1}$, is its effective Compton length. We have that 

\bea \lambda_{\phi_*}^\text{atm}\approx 10^{18}\lambda_\phi^\text{atm},\label{lambda-ratio}\eea where $\lambda_\phi^\text{atm}\sim 1$mm, is the Compton length of the chameleon field which is consistent with the experiments on fifth-force, while $\lambda_{\phi_*}^\text{atm}$ is the effective range of the scalar field mediated interaction, computed with the potential $V(\phi)\propto\phi^4$, under the assumption that the cosmological bound $\lambda_{\phi_*}^\text{cosm}\sim 10^{26}$m, is met: $$\lambda^\text{atm}_{\phi_*}\sim 10^{-11}\lambda^\text{cosm}_{\phi_*}\sim 10^{15}\,\text{m}=10^{18}\,\text{mm}.$$ Then, requiring that $$\frac{\Delta U_N^*}{\Delta U_N}=\left(\frac{\beta^*_\text{eff}}{\beta_\text{eff}}\right)^2\frac{e^{r/\lambda_\phi^\text{atm}}}{e^{r/\lambda_{\phi_*}^\text{atm}}}\approx 1,$$ for the given potential $V(\phi)\propto\phi^4$, the expected weakening of the effective coupling of the chameleon BD field to the surrounding matter, is an unnaturally large effect: $\left(\beta^*_\text{eff}\right)^2\sim\exp\left(-10^{20}\right)\beta_\text{eff}^2,$ where we have assumed that $\beta_\text{eff}\sim 1$. In order to obtain the above estimate, we have arbitrarily set the distance from the source of gravity $r\approx 10^2\lambda_{\phi_*}^\text{atm}$, and the Eq. (\ref{lambda-ratio}) has been considered.

The above results are true, in general, for power--law potentials of arbitrary power: $V(\phi)\propto\phi^\lambda$. In this latter case, for the mass squared of the BD field, one gets: $m^2\propto\rho^{(\lambda-1)/\lambda}$, where, as $\lambda\rightarrow\infty$, $(\lambda-1)/\lambda\rightarrow 1$. This means that the latter power can never exceed unity $(\lambda-1)/\lambda\leq 1$. Recall that, a necessary requirement, when the power--law potential $V(\phi)\propto\phi^\lambda$ is allowed to explain cosmological and terrestrial bounds at once, amounts to: $(\lambda-1)/\lambda>2.071$. 

Our conclusion is that, in general, terrestrial and solar system bounds on the mass of the BD scalar field, and bounds of cosmological origin, are difficult to reconcile through a single chameleon potential.


\section{the chameleonic mass of the dilaton}\label{dilaton-mass-sec}

In several situations of interest, in particular when the focus is in the low--energy effective string theory, it is useful to turn to the field variables $\vphi$, and $U=U(\vphi)$, in Eq. (\ref{vphi}). This choice singles out the string frame formulation of this theory. In terms of these variables the effective mass squared of the dilaton, i. e., the one which is obtained from Eq. (\ref{bd-mass}) by substitution of (\ref{vphi}), is written as:

\bea m_\vphi^2(\vphi)=\frac{2}{3+2\omega_\textsc{bd}}\left[\der^2_\vphi U(\vphi)-U(\vphi)\right].\label{dilaton-mass}\eea  However, as we will immediately show, the above expression for the effective mass of the dilaton, has to be taken with caution. 

If follow, step by step, the procedure in sections \ref{bd-mass-sec} and \ref{cham-mass-sec}, then, the KG equation (\ref{dilaton-kg}) can be rewritten as:

\bea \nabla^2\vphi+(\der\vphi)^2=\der_\vphi U_\text{ch},\label{dilaton-kg-eq}\eea where the dilatonic chameleon potential $U_\text{ch}=U_\text{ch}(\vphi)$, is defined as: 

\bea U_\text{ch}=\frac{2}{3+2\omega_\textsc{bd}}\left[U-\int d\vphi U-\frac{1}{2}\,e^{-\vphi}T^{(m)}\right],\label{dilaton-cham-pot}\eea so that $$\der_\vphi U_\text{ch}=\frac{2}{3+2\omega_\textsc{bd}}\left[\der_\vphi U-U+\frac{1}{2}\,e^{-\vphi}T^{(m) }\right],$$ coincides with the RHS of Eq. (\ref{dilaton-kg}).

Assuming that $U_\text{ch}$ is a minimum at some $\vphi_*$, small deviations about $\vphi_*$: $\vphi=\vphi_*+\delta\vphi$ ($\delta\vphi\ll 1$), $\der_\mu\delta\vphi\rightarrow\der_\mu\vphi$ $\Rightarrow\;(\der\delta\vphi)^2\sim O(\delta\vphi^2)$, so that, the second term in the left-hand side (LHS) of Eq. (\ref{dilaton-kg-eq}), may be ignored, and one ends up with the standard KG equation for the fluctuations of the dilaton: $\nabla^2\delta\vphi=\der_\vphi U_\text{ch}$. Since $U_\text{ch}$ develops a minimum at $\vphi_*$, then, one may define the effective mass (squared) of the dilaton: $m_{\vphi_*}^2=\der^2_\vphi U_\text{ch}(\vphi_*)$, which is the one that determines the range of the Yukawa interaction in the string frame: $$\delta\vphi(r)\propto \frac{e^{-m_{\vphi_*} r}}{r}.$$

Following the procedure exposed in the former sections \ref{bd-mass-sec} and \ref{cham-mass-sec}, a straightforward generalization $$m_\vphi^2(\vphi)=\der^2_\vphi U_\text{ch}(\vphi)\;\rightarrow\;m_{\vphi_*}^2=\der^2_\vphi U_\text{ch}(\vphi_*),$$ leads to the following definition of the effective field-theoretical mass of the dilaton in the string frame:

\bea m_\vphi^2(\vphi)=\frac{2}{3+2\omega_\textsc{bd}}\left[\der^2_\vphi U-\der_\vphi U-\frac{1}{2}\,e^{-\vphi}T^{(m)}\right].\label{eff-dilaton-mass}\eea 

It is not Eq. (\ref{dilaton-mass}), but the latter equation (\ref{eff-dilaton-mass}), the one to be contrasted with the effective field--theoretical mass parameter (\ref{bd-mass}) of the BD scalar field in the Jordan frame. Nevertheless, since, as assumed, $U_\text{ch}$ develops a minimum at some $\vphi_*$, then 

\bea &&\der_\vphi U_\text{ch}(\vphi_*)=0\;\Rightarrow\nonumber\\
&&\der_\vphi U(\vphi_*)=U(\vphi_*)-\frac{1}{2}\,e^{-\vphi_*}T^{(m)},\nonumber\eea so that,

\bea &&\der_\vphi^2 U_\text{ch}(\vphi_*)=\nonumber\\
&&\frac{2}{3+2\omega_\textsc{bd}}\left[\der_\vphi^2 U(\vphi_*)-\der_\vphi U(\vphi_*)-\frac{e^{-\vphi_*}}{2}\,T^{(m)}\right]\nonumber\\
&&\;\;\;\;\;\;\;\;\;\;\;\;\;\;\;\;\;=\frac{2}{3+2\omega_\textsc{bd}}\left[\der_\vphi^2 U(\vphi_*)-U(\vphi_*)\right],\label{compare}\eea where the last equation above coincides with Eq. (\ref{dilaton-mass}), when given quantities are evaluated at the minimum $\vphi_*$ of the chameleon potential (\ref{dilaton-cham-pot}).

\subsection{Chameleonic dilaton}\label{cham-mass-subsec}

For sake of illustration, let us assume, as before, a pressureless, homogeneous and isotropic background with energy density $\rho$. Consider an exponential potential

\bea U(\vphi)=M^2\,e^{\lambda\vphi},\label{exp-pot}\eea where $M$ and $\lambda$ are free constants. Notice that, under the replacements $\phi\rightarrow e^\vphi$, $M^{4-n}\rightarrow M^2$, and $n-1\rightarrow\lambda$, the latter potential in Eq. (\ref{exp-pot}) is mapped into the potential $$V(\phi)=M^{4-n}\phi^n,$$ in terms of the JF--BD scalar field $\phi$. The particular value $\lambda=3$ ($n=4$), corresponds to the quartic potential discussed above. According to Eq. (\ref{dilaton-cham-pot}), the resulting chameleon potential looks like:

\bea U_\text{ch}(\vphi)=\frac{2M^2(\lambda-1)}{\lambda(3+2\omega_\textsc{bd})}\left[e^{\lambda\vphi}+\frac{\lambda\rho\;e^{-\vphi}}{2M^2(\lambda-1)}\right],\label{exp-cham-pot}\eea where, as before, we have assumed a homogeneous pressureless fluid with trace of the stress-energy tensor $T^{(m)}=-\rho$. 

For $\lambda>1$, the above potential is a minimum at\footnote{For $\lambda<1$, the chameleon potential (\ref{exp-cham-pot}) does not develop minimums at all.} 

\bea \vphi_*=\ln\left[\frac{\rho}{2M^2(\lambda-1)}\right]^\frac{1}{\lambda+1},\label{sf-min}\eea where the effective mass squared of the fluctuations of the chameleonic dilaton is given by $m^2_{\vphi_*}=\der_\vphi^2 U_\text{ch}(\vphi_*)$:

\bea m^2_{\vphi_*}=\frac{(\lambda+1)\left[2M^2(\lambda-1)\right]^\frac{1}{\lambda+1}}{3+2\omega_\textsc{bd}}\,\rho^\frac{\lambda}{\lambda+1}.\label{exp-cham-mass}\eea  

For the particular value $\lambda=3$ which, as mentioned, corresponds to the quartic potential in the Jordan frame of BD theory: $V(\phi)\propto\phi^4$, the chameleonic mass of the dilaton goes like $m_{\vphi_*}\propto\rho^{3/8}$, which is consistent with the results of section \ref{estimate-sec}, as one should expect. 

As explained in former sections, an interesting thing about the mass of the dilaton is related with the fact that, for the specific exponential potential $U(\vphi)\propto e^\vphi$, which, in terms of the BD field $\phi$, corresponds to the quadratic monomial potential $V(\phi)\propto\phi^2$, the chameleonic dilaton is massless. Actually, in this case, in Eq. (\ref{exp-pot}) one have to set $\lambda=1$. When this value of the free parameter is substituted into Eq. (\ref{exp-cham-mass}) one gets $m^2_{\vphi_*}=0$. This conclusion is correct even if consider the effective potential (\ref{eff-pot}). As shown in section \ref{q-pot-subsec}, in the Jordan frame of the BD theory, where the effective mass squared of the BD field is given by Eq. (\ref{bd-mass}), for the quadratic monomial $V(\phi)=M^2\phi^2/2$, which corresponds to the exponential potential for the dilaton $U(\vphi)=M^2\exp(\vphi)/2$, since $\der_\phi V=M^2\phi$, and $\der^2_\phi V=M^2$, the mass squared (\ref{bd-mass}) vanishes: $$m_\phi^2=\frac{2}{3+2\omega_\textsc{bd}}\left[\phi \der^2_\phi V(\phi)-\der_\phi V(\phi)\right]=0.$$ Hence, the potential $$U(\vphi)\propto e^\vphi\;\Leftrightarrow\;V(\phi)\propto\phi^2,$$ does not provides an effective mass for the BD scalar field. 

This means that, for the exponential potential (\ref{exp-pot}) with the specific value of the free parameter $\lambda=1$ (or for the quadratic monomial if working with the Jordan frame BD variables), the chameleon effect does not work, and the dilaton can not be screened from solar systems experiments. This looks like a bad news for the dilaton since, as shown in Ref. \cite{hrycyna, bd-quiros} (see the next section), only for the exponential potential $U(\vphi)=M^2\exp\vphi$, or for potentials which asymptote to $M^2\exp\vphi$, the BD theory has the $\Lambda$CDM solution -- also called as concordance model \cite{lcdm, c-c-ratra-peebles} -- as a global attractor.


\section{what do the dynamical systems have to say about the brans--dicke chameleon?}\label{dyn-syst-sec}

In this section we shall consider Friedmann-Robertson-Walker (FRW) spacetimes with flat spatial sections for which the line-element takes the simple form: $$ds^2=-dt^2+a^2(t)\delta_{ij}dx^idx^j,\;i,j=1,2,3\,.$$ We assume the matter content of the Universe in the form of a cosmological perfect fluid, which is characterized by the following state equation $p_m=w_m\rho_m$, relating the barotropic pressure $p_m$ and the energy density $\rho_m$ of the fluid, where $w_m$ is the so called equation of state (EOS) parameter. Under these assumptions the cosmological equations which are derived from (\ref{dilaton-action}), are written as it follows:

\bea &&3H^2=\frac{\omega_\textsc{bd}}{2}\,\dot\vphi^2-3H\dot\vphi+U+e^{-\vphi}\rho_m,\nonumber\\
&&\dot H=-\frac{\omega_\textsc{bd}}{2}\,\dot\vphi^2+2H\dot\vphi+\frac{\der_\vphi U-U}{3+2\omega_\textsc{bd}}\nonumber\\
&&\;\;\;\;\;\;\;\;\;\;\;\;\;\;\;-\frac{2+\omega_\textsc{bd}\left(1+w_m\right)}{3+2\omega_\textsc{bd}}\,e^{-\vphi}\rho_m,\nonumber\\
&&\ddot\vphi+3H\dot\vphi+\dot\vphi^2=2\frac{U-\der_\vphi U}{3+2\omega_\textsc{bd}}+\frac{1-3w_m}{3+2\omega_\textsc{bd}}\,e^{-\vphi}\rho_m,\nonumber\\
&&\dot\rho_m+3H\left(w_m+1\right)\rho_m=0,\label{efe}\eea where $H\equiv\dot a/a$ is the Hubble parameter.  

Due to the complexity of the system of non-linear second-order differential equations (\ref{efe}), it is a very difficult (and perhaps unsuccessful) task to find exact solutions. Even when an analytic solution can be found it will not be unique but just one in a large set of them. This is in addition to the problem of the stability of given solutions. In this case the dynamical systems tools come to our rescue. The dynamical systems theory provides powerful tools which are commonly used in cosmology to extract essential information on the dynamical properties of a variety of cosmological models, in particular, those models where the scalar field plays a role \cite{wands, coley, copeland-rev, copeland, quiros-prd-2009, luis-mayra, luis, bohmer-rev, epj-2015, genly, cosmology-books, amendola, uggla-ref, ds-bd, hrycyna, holden, olga, faraoni, iranies, indios, genly-1}. 

In general, when one deals with BD cosmological models it is customary to choose the following variables \cite{hrycyna}:

\bea x\equiv\frac{\dot\vphi}{\sqrt{6}H}=\frac{\vphi'}{\sqrt 6},\;y\equiv\frac{\sqrt{U}}{\sqrt{3}H},\;\xi\equiv 1-\frac{\der_\vphi U}{U},\label{vars}\eea where the tilde means derivative with respect to the variable $\tau\equiv\ln a$, i. e., to the number of e-foldings. As a matter of fact $x$ and $y$ in Eq. (\ref{vars}), are the same variables which are usually considered in similar dynamical systems studies of FRW cosmology, within the frame of Einstein's general relativity with a scalar field matter source \cite{wands}. In terms of the above variables the Friedmann constraint can be written as \cite{bd-quiros}

\bea \Omega^\text{eff}_m\equiv\frac{e^{-\vphi}\rho_m}{3H^2}=1+\sqrt{6}x-\omega_\textsc{bd}\,x^2-y^2\geq 0.\label{friedmann-c}\eea 

Notice that one might define a dimensionless potential energy density and an ``effective kinetic'' energy density

\bea \Omega_U=\frac{U}{3H^2}=y^2,\;\Omega^\text{eff}_K=x\left(\omega_\textsc{bd}x-\sqrt{6}\right),\label{omega-u-k}\eea respectively, so that the Friedmann constraint can be re-written in the following compact form: $$\Omega^\text{eff}_K+\Omega_U+\Omega^\text{eff}_m=1.$$ 

The definition for the dimensionless effective kinetic energy density $\Omega^\text{eff}_K$ has not the same meaning as in GR with a scalar field. It may be a negative quantity without challenging the known laws of physics. Besides, since there is not restriction on the sign of $\Omega^\text{eff}_K$, then, it might happen that $\Omega_U=U/3H^2>1$. This is due to the fact that the dilaton field in the BD theory is not a standard matter field but it is a part of the gravitational field itself. This effective (dimensionless) kinetic energy density vanishes whenever: $$x=\frac{\sqrt{6}}{\omega_\textsc{bd}}\;\Rightarrow\;\dot\vphi=\frac{6}{\omega_\textsc{bd}}\,H\;\Rightarrow\;\vphi=\frac{6}{\omega_\textsc{bd}}\,\ln a,$$ or if: $x=0\;\Rightarrow\;\dot\vphi=0\;\Rightarrow\;\vphi=const.,$ which, provided that the matter fluid is cold dark matter, corresponds to the GR--de Sitter universe, i. e., to the $\Lambda$CDM model.\footnote{It is known that, but for some anomalies in the power spectrum of the cosmic microwave background \cite{planck-2013}, at the present stage of the cosmic evolution, any cosmological model has to approach to the so called concordance or $\Lambda$CDM model \cite{lcdm}. The mathematical basis for the latter is the Einstein--Hilbert action plus a matter action piece: $$S_{\Lambda\textsc{cdm}}=\frac{1}{2}\int d^4x\sqrt{|g|}\left(R-2\Lambda\right)+\int d^4x\sqrt{|g|}{\cal L}_\textsc{cdm},$$ where ${\cal L}_\textsc{cdm}$ is the Lagrangian density of (pressureless) cold dark matter (CDM). The $\Lambda$CDM action above can be obtained from (\ref{dilaton-action}), provided that the dilaton acquires some expectation value $\vphi_0$, so that $$e^{\vphi_0}=\frac{1}{2},\;U(\vphi_0)=\Lambda,\;{\cal L}_\textsc{cdm}\rightarrow\,{\cal L}_m.$$}

\subsection{The dynamical system}

Our goal will be to write the resulting system of cosmological equations (\ref{efe}), in the form of a system of autonomous ordinary differential equations (ODE-s) in terms of the variables $x$, $y$, $\xi$, of some phase space. We obtain the following autonomous system of ODE-s:

\bea &&x'=-3x\left(1+\sqrt{6}x-\omega_\textsc{bd}x^2\right)+\frac{x+\sqrt{2/3}}{3+2\omega_\textsc{bd}}\,3y^2\xi\nonumber\\
&&\;\;\;\;\;\;\;\;\;\;\;\;\;\;+\frac{\frac{1-3w_m}{\sqrt 6}+\left[2+\omega_\textsc{bd}(1+w_m)\right]\,x}{3+2\omega_\textsc{bd}}\,3\Omega^\text{eff}_m,\nonumber\\
&&y'=y\left[3x\left(\omega_\textsc{bd}x-\frac{\xi+3}{\sqrt{6}}\right)+\frac{3y^2\xi}{3+2\omega_\textsc{bd}}\right.\nonumber\\
&&\left.\;\;\;\;\;\;\;\;\;\;\;\;\;\;\;\;\;\;\;\;\;\;\;\;\;\;\;\;+\frac{2+\omega_\textsc{bd}\left(1+w_m\right)}{3+2\omega_\textsc{bd}}\,3\Omega^\text{eff}_m\right],\nonumber\\
&&\xi'=-\sqrt{6}x\left(1-\xi\right)^2\left(\Gamma-1\right),\label{asode}\eea where $\Omega^\text{eff}_m$ is given by Eq. (\ref{friedmann-c}), and it is assumed that $\Gamma=U\der^2_\vphi U/(\der_\vphi U)^2$ can be written as a function of $\xi$ \cite{epj-2015}: $\Gamma=\Gamma(\xi)$. Hence, the properties of the dynamical system (\ref{asode}) are highly dependent on the specific functional form of the potential $U=U(\vphi)$.

\subsection{Vacuum Brans--Dicke cosmology}

A significant simplification of the dynamical equations is achieved when matter degrees of freedom are not considered. In this case, since $\Omega^\text{eff}_m=0\;\Rightarrow\;y^2=1+\sqrt{6}x-\omega_\textsc{bd}\,x^2,$ then the system of ODE-s (\ref{asode}) simplifies to a plane-autonomous system of ODE-s:

\bea &&x'=\left(-3x+3\frac{x+\sqrt{2/3}}{3+2\omega_\textsc{bd}}\,\xi\right)\left(1+\sqrt{6}x-\omega_\textsc{bd}x^2\right),\nonumber\\
&&\xi'=-\sqrt{6}x\left(1-\xi\right)^2\left(\Gamma-1\right).\label{x-xi-ode-vac}\eea 

In the present case one has

\bea &&\Omega_U=\frac{U}{3H^2}=y^2=1+\sqrt{6}x-\omega_\textsc{bd}x^2,\nonumber\\
&&\Omega^\text{eff}_K=x\left(\omega_\textsc{bd}x-\sqrt{6}\right)\;\Rightarrow\;\Omega^\text{eff}_K+\Omega_U=1,\label{omegas-vac}\eea where we recall that the definition of the effective (dimensionless) kinetic energy density $\Omega^\text{eff}_K$, has not the same meaning as in GR with scalar field matter, and it may be, even, a negative quantity. Here we consider non-negative self-interaction potentials $U(\vphi)\geq 0$, so that the dimensionless potential energy density $\Omega_U=y^2$, is restricted to be always non-negative: $\Omega_U=1+\sqrt{6}x-\omega_\textsc{bd}x^2\geq 0$. Otherwise, $y^2<0$, and the phase-plane would be a complex plane. Besides, we shall be interested in expanding cosmological solutions exclusively ($H\geq 0$), so that $y\geq 0$. Because of this the variable $x$ is bounded to take values within the following interval:

\bea \alpha_-\leq x\leq\alpha_+,\;\alpha_\pm=\sqrt\frac{3}{2}\left(\frac{1\pm\sqrt{1+2\omega_\textsc{bd}/3}}{\omega_\textsc{bd}}\right).\label{x-bound}\eea This means that the phase space for the vacuum Brans--Dicke theory $\Psi_\text{vac}$ can be defined as follows:

\bea &&\Psi_\text{vac}=\left\{(x,\xi):\;\alpha_-\leq x\leq\alpha_+\right\},\label{vac-phase-space}\eea where the bounds on the variable $\xi$ --  if any -- are set by the concrete form of the self-interaction potential.

There are found four dilatonic equilibrium points $P_i:(x_i,\xi_i)$, in the phase space corresponding to the dynamical system (\ref{x-xi-ode-vac}). One of them is the GR--de Sitter phase: 

\bea &&P_\text{dS}:(0,0)\;\Rightarrow\;x=0\;\Rightarrow\;\vphi=\vphi_0,\;\text{and}\nonumber\\
&&\;\;\;\;\;\;\;\;\;\;\;\;\;\;\;y^2=1\;\Rightarrow\;3H^2=U=const.,\nonumber\eea which, since $\dot H$ given by 

\bea &&\frac{\dot H}{H^2}=\left(2\sqrt{6}-3\omega_\textsc{bd}\,x\right)x\nonumber\\
&&\;\;\;\;\;\;\;\;\;\;\;\;\;\;\;\;\;\;\;\;\;\;\;\;-\frac{3\left(1+\sqrt{6} x-\omega_\textsc{bd} x^2\right)\xi}{3+2\omega_\textsc{bd}},\label{h-dot}\eea vanishes, then $\dot H=0$ $\Rightarrow\;H=H_0$, corresponds to accelerated expansion $q=-1-\dot H/H^2=-1.$ Besides, $$\xi=0\;\Leftrightarrow\;\frac{\der_\vphi U}{U}=1\;\Rightarrow\;U(\vphi)\propto e^\vphi,$$ i. e., this point exists for this specific exponential potential exclusively. The eigenvalues of the linearization matrix around $(0,0)$ are: $\lambda_1=-3,\;\lambda_2=0.$ This means that $(0,0)$ is a non-hyperbolic point.

We found, also, another de Sitter solution: $q=-1$ $\Rightarrow\;\dot H=0$, which is associated with scaling of the effective kinetic and potential energies of the dilaton: $$P^\textsc{bd}_\text{dS}:\left(\frac{1}{\sqrt{6}(1+\omega_\textsc{bd})},1\right)\;\Rightarrow\;\frac{\Omega^\text{eff}_K}{\Omega_U}=-\frac{6+5\omega_\textsc{bd}}{12+17\omega_\textsc{bd}+6\omega^2_\textsc{bd}},$$ $\lambda_1=-(4+3\omega_\textsc{bd})/(1+\omega_\textsc{bd})$, $\lambda_2=0,$ where, as before, $\lambda_1$ and $\lambda_2$ are the eigenvalues of the linearization matrix around the critical point. We call this as BD--de Sitter critical point to differentiate it from the GR--de Sitter point.

In order to make clear what the difference is between both de Sitter solutions,\footnote{The existence of the BD--de Sitter solution can be traced back to the work in the second item of Ref. \cite{barrow-ref}.} let us note that the Friedmann constraint (\ref{friedmann-c}), evaluated at the BD--de Sitter point above, can be written as $$e^{-\vphi}\rho_m=3H_0^2+\frac{6+5\omega_\textsc{bd}}{6(1+\omega_\textsc{bd})^2}\,3H_0^2-U_0,$$ i. e., $e^{-\vphi}\rho_m=const.$ This means that the weakening/strengthening of the effective gravitational coupling ($G_\text{eff}\propto e^{-\vphi}$) is accompanied by a compensating growing/decreasing property of the energy density of matter $\rho_m\propto e^\vphi$, which leads to an exponential rate of expansion $a(t)\propto e^{H_0 t}$. This is to be contrasted with the GR--de Sitter solution: $3H_0^2=U_0$ $\Rightarrow\;a(t)\propto e^{\sqrt{U_0/3}\,t}$, which is obtained only for vacuum, $\rho_\text{vac}=U_0$; $\rho_m=0$. 

The effective stiff-dilaton critical points ($\Omega^\text{eff}_K=1$): $$P_\pm:\left(\alpha_\pm,1\right)\;\Rightarrow\;q_\pm=2+\sqrt{6}\,\alpha_\pm,$$ $$\lambda^\pm_1=6\left(1+\sqrt\frac{2}{3}\,\alpha_\pm\right),\;\lambda_2=0,$$ are also found, where the $\alpha_\pm$ are defined in Eq. (\ref{x-bound}).

\subsection{The exponential potential}

In the general case when we have the exponential potential (\ref{exp-pot}), since $\xi=1-\lambda$, is a constant, i. e., it can not be a phase space variable anymore, the plane-autonomous system of ODE-s (\ref{x-xi-ode-vac}), simplifies to a single autonomous ODE:

\bea &&x'=-\left[\frac{\left(\lambda+2+2\omega_\textsc{bd}\right)x-\sqrt\frac{2}{3}(1-\lambda)}{1+2\omega_\textsc{bd}/3}\right]\times\nonumber\\
&&\;\;\;\;\;\;\;\;\;\;\;\;\;\;\;\;\;\;\;\;\;\;\;\;\;\;\;\;\;\;\;\;\times\left(1+\sqrt{6}x-\omega_\textsc{bd}x^2\right).\label{x-ode-vac-exp}\eea 

The critical points of the latter dynamical system are:

\bea x_1=\frac{\sqrt{2/3}\,(1-\lambda)}{\lambda+2+2\omega_\textsc{bd}},\;x_\pm=\alpha_\pm,\label{vac-exp-c-points}\eea where the $\alpha_\pm$ are given by Eq. (\ref{x-bound}). Notice that, since $x_i\neq 0$ (but for $k=1$, in which case $x_1=0$ and $q=-1$), there are not critical points associated with constant $\vphi=\vphi_0$. This means that the de Sitter phase with $\dot\vphi=0$ ($\vphi=const$), $U(\vphi)=const.$, i. e., the one which occurs in GR and which stands at the heart of the $\Lambda$CDM model, does not arise in the general case when $\lambda\neq 1$. 

Hence, only in the particular case of the exponential potential (\ref{exp-pot}) with $\lambda=1$ ($\xi=0$), which corresponds to the quadratic potential in terms of the JF--BD variables: $V(\phi)=M^2\phi^2$, the GR--de Sitter phase is a critical point of the dynamical system (\ref{x-ode-vac-exp}). In this case the critical points are (see Eq. (\ref{vac-exp-c-points})): $x_1=0$, $x_\pm=\alpha_\pm$. Worth noticing that $x_1=0$ corresponds to the GR--de Sitter solution $3H^2=M^2\exp\vphi_0$, meanwhile, the $x_\pm=\alpha_\pm$, correspond to the stiff-fluid (kinetic energy) dominated phase: $\Omega^\text{eff}_K=1$. 

For small (linear) perturbations $\epsilon=\epsilon(\tau)$ around the GR--de Sitter critical point: $x=0+\epsilon$, $\epsilon\ll 1$, one has that: $\epsilon'=-3\epsilon$ $\Rightarrow\;\epsilon(\tau)\propto\exp(-3\tau)$, so that it is an attractor solution, independent on the value of the BD coupling constant $\omega_\textsc{bd}$.\footnote{Around the stiff-matter solutions $x=\alpha_\pm+\epsilon_\pm$: $$\epsilon_\pm(\tau)\propto e^{3\left(2+\sqrt{6}\,\alpha_\pm\right)\tau},$$ so that, if assume non-negative $\omega_\textsc{bd}\geq 0$, the points $x_\pm$ are always past attractors (unstable equilibrium points) since $2+\sqrt{6}\,\alpha_->0$. For negative $\omega_\textsc{bd}<0$, these points are both past attractors whenever $\omega_\textsc{bd}<-3/2$. In this latter case, for $-3/2<\omega_\textsc{bd}<0$, the point $x_+$ is a past attractor, while the point $x_-$ is a future attractor instead.}

In \cite{bd-quiros} it has been shown that, in general, only for the specific exponential potential $U(\vphi)=M^2\exp\vphi$, or for potentials which asymptote to $M^2\exp\vphi$, the $\Lambda$CDM model, which is associated with the GR--de Sitter point, can be an attractor of the Brans--Dicke theory. In other words, only BD cosmological models with $U(\vphi)=M^2\exp\vphi$, or $U(\vphi)\rightarrow M^2\exp\vphi$, can pass the present observational cosmological tests.


\section{emergence of the GR--de Sitter local attractor}\label{gr-attractor-sec}

It is essential for the chameleon mechanism to work, that the chameleon potential $V_\text{ch}(\phi)$ given by Eq. (\ref{bd-cham-pot}), be a minimum at some $\phi_*$. Once this condition is met and the chameleon mechanism is in action, the BD field does oscillations around $\phi_*$, and it acquires an effective chameleonic mass as explained (see section \ref{perts-subsec}). This is more easily seen in a cosmological context where, due to the expansion of the universe, the KG equation for the chameleonic BD field develops a friction term $\propto\dot\phi$ which causes the oscillations to damp.\footnote{One important example is the BD--KG equation in a cosmological Friedmann-Robertson-Walker space.} In consequence, the BD chameleon field does damped oscillations until it settles down in the minimum of the potential. Once this stable state is reached, the BD theory transmutes into general relativity with a cosmological $\Lambda$--term. 

The latter conclusion is true, however, only in the case when the density of the surrounding matter is a constant. Otherwise, if $\rho=\rho(t)$ were a function of the cosmic time, we would not have a point of minimum of the chameleon potential, but a whole curve: $\vphi_*=\vphi_*(\rho(t))$. This would entail that $$\dot\vphi_*(t)=\frac{\der\vphi_*}{\der\rho}\,\dot\rho(t)\neq 0,$$ the $t$--gradient of the dilaton will be aligned with the $t$--gradient of the density profile, so that we would not get general relativity but Brans--Dicke theory will be retained instead.

\subsection{Transmutation of BD theory into GR}

In order to visualize how the transmutation of BD theory into GR--de Sitter theory operates, let us take a look at the Jordan frame BD action (\ref{bd-action}). We will assume that the given effective chameleon potential is a minimum at some constant expectation value of the JF--BD field $\phi=\phi_*$, within a region of constant matter density. As commented above, these are necessary requirements for the transmutation of BD theory into general relativity. 

At the minimum $\phi_*$, the gravitational sector of (\ref{bd-action}) transforms into the Einstein--Hilbert action

\bea S^*=\int d^4x\sqrt{|g|}\,\phi_*\left[R-2\frac{V(\phi_*)}{\phi_*}\right].\label{eh-action}\eea 

In a similar way, at $\phi=\phi_*$, the BD field equations (\ref{bd-feq}), (\ref{bd-kg}), transform into the Einstein's field equations of general relativity with a cosmological constant: $$G_{\mu\nu}=\frac{1}{\phi_*}\,T^{(m)}_{\mu\nu}-\frac{V(\phi_*)}{\phi_*}\,g_{\mu\nu},$$ where, in the above equations, we have to set $$\phi_*=M^2_\textsc{pl}=\frac{1}{8\pi G},\;\text{and}\;\frac{V(\phi_*)}{\phi_*}=\Lambda,$$ respectively. Hence, the chameleon effect in the BD theory warrants that the $\Lambda$CDM model is a local attractor of the Brans--Dicke cosmology. We have to point out, once again, that for this to happen it is indispensable that the given chameleon potential develops a minimum in a region of constant density. Otherwise the former statement would not be true. This means the the GR--de Sitter solution can be, at most, a local attractor (it can be also a saddle point), since a global attractor would entail a constant density of the cosmological background. But, as known, any form of matter in a cosmological setting, but for the quantum vacuum with constant density $\rho_\text{vac}$, would have, necessarily, an evolving density $\rho=\rho(t)$ (see subsection \ref{damped-subsec}).

In order to compare with the results of the dynamical systems study (see section \ref{dyn-syst-sec}), it is easier to work with the dilatonic variable. Hence, assuming that the chameleon potential $U_\text{ch}(\vphi)$ in Eq. (\ref{dilaton-cham-pot}), is a minimum at some constant $\vphi=\vphi_*$, within a region of constant density $\rho$, at the minimum the SF--BD equation of motion (\ref{dilaton-feq}) decays into the Einstein's GR equation: $$G_{\mu\nu}=e^{-\vphi_*}T^{(m)}_{\mu\nu}-g_{\mu\nu} U(\vphi_*),$$ where $e^{\vphi_*}=M^2_\textsc{pl}$, and $\Lambda=U(\vphi_*)$. In terms of the FRW metric we can say that the cosmological SF--BD equations (\ref{efe}), transmute into the standard cosmological equations:

\bea &&3H^2=e^{-\vphi_*}\,\rho_m+U(\vphi_*)=M^2_\textsc{pl}\,\rho_m+\Lambda,\nonumber\\
&&\dot H=-\frac{w_m+1}{2}\,e^{-\vphi_*}\,\rho_m=-\frac{w_m+1}{2}\,M^2_\textsc{pl}\,\rho_m.\nonumber\eea

Let us assume, for definiteness, that the spacetime dynamics is driven by the exponential potential (\ref{exp-pot}): $U(\vphi)=M^2\exp(\lambda\vphi)$. Then, as shown, for $\lambda>1$, the corresponding chameleon potential (\ref{exp-cham-pot}) is a minimum at (see equation (\ref{sf-min})): $$\vphi_*=\ln\left[\frac{\rho}{2M^2(\lambda-1)}\right]^\frac{1}{\lambda+1}.$$ 

Due to the expansion, i. e., to the friction term in the FRW--KG equation $\propto H\dot\vphi$, within the region with constant $\rho$ the dilaton does damped oscillations around this $\vphi_*$ until, eventually, the SF--BD theory gives way to Einstein's GR with a cosmological constant. This result is quite independent of the assumed dilaton potential $U(\vphi)$. In the discussed case (the exponential potential $U(\vphi)\propto\exp\vphi$), for instance, the result is independent of $\lambda$ ($\lambda>1$). 

How to reconcile this result with the result of section \ref{dyn-syst-sec} (see Ref. \cite{hrycyna, bd-quiros})? In section \ref{dyn-syst-sec}, by means of the tools of the dynamical systems theory, it has been shown that, only for the exponential potential $U(\vphi)=M^2\exp\vphi$ (this corresponds to the quadratic potential $V(\phi)\propto\phi^2$ in terms of the JF--BD field), or for potentials that asymptote to $M^2\exp\vphi$, the GR--de Sitter solution is an attractor of the BD theory. Meanwhile, in this subsection it has been shown that the GR--de Sitter space is an attractor of SF--BD theory, provided that the effective chameleon potential is a minimum within a region of constant density of matter. This result has been shown for the most general exponential potential $U(\vphi)\propto\exp(\lambda\vphi)$, independent of the free parameter $\lambda$. I. e., the $\Lambda$CDM model is an atractor of SF--BD theory not only for the specific exponential potential $U(\vphi)\propto\exp\vphi$, but for arbitrary exponentials.

The answer to the above question is simple enough: in a cosmological setting, but for the quantum vacuum, the density of the background matter is an evolving function of the cosmic time $t$: $\rho=\rho(t)$. Hence, assuming that the chameleon potential $U_\text{ch}(\vphi)$ is a minimum at some $$\vphi_*=\vphi_*(\rho(t))\;\Rightarrow\;\dot\vphi_*=\frac{d\vphi_*}{d\rho}\,\dot\rho\neq 0,$$ which means, in turn, that at the minimum of the chameleon potential what we have is Brans--Dicke theory and not GR. Besides, since $\vphi_*=\vphi_*(\rho(t))$ is an implicit function of the cosmic time, then the resulting effective mass squared $m^2_{\vphi_*}=m^2_{\vphi_*}(\rho(t))$, would be so. Consequently, this can not be a stable state of BD theory, since in the latter the masses of particles (including any field excitations) are constants. This entails that the GR--de Sitter solution can be, at most, a saddle point, unless either the potential $U(\vphi)\propto\exp\vphi$, or it asymptotes to the exponential $U(\vphi)\rightarrow\exp\vphi$, as it has been shown in section \ref{dyn-syst-sec}. In general, only for regions of constant density of matter, for instance, regions filled with quantum vacuum, the GR--de Sitter space can be a (local) attractor of the BD theory.

\subsection{Damped oscillations of the dilaton}\label{damped-subsec}

In order to illustrate our arguments above, let us investigate the oscillations of the dilaton around the minimum of the exponential potential in a FRW spacetime with flat spatial sections (see section \ref{dyn-syst-sec}). As before we shall assume that the FRW spacetime is filled with pressureless dust. The corresponding SF--KG equation (\ref{dilaton-kg}) reads:

\bea &&\ddot\vphi+3H\dot\vphi+\dot\vphi^2=\nonumber\\
&&\;\;\;\;\;\;\;\;\;\;\;\;\;-\frac{2M^2(\lambda-1)}{3+2\omega_\textsc{bd}}\left[e^{\lambda\vphi}-\frac{\rho_m\,e^{-\vphi}}{2M^2(\lambda-1)}\right].\label{frw-dilaton-kg}\eea The second term in the LHS of this equation, i. e., the friction term $\propto H\dot\vphi$, is due to the implicit consideration of the curvature effects in FRW cosmological backgrounds.

Next we expand this equation around the minimum of the chameleon potential (\ref{exp-cham-pot}): 

\bea &&e^{\lambda\vphi_*}-\frac{\rho_m\,e^{-\vphi_*}}{2M^2(\lambda-1)}=0\nonumber\\
&&\;\;\;\;\;\;\;\;\;\Rightarrow\;\vphi_*=\ln\left[\frac{\rho_m}{2M^2(\lambda-1)}\right]^\frac{1}{\lambda+1}.\label{min-cham-exp-pot}\eea Hence, in Eq. (\ref{frw-dilaton-kg}), we make the replacement: $$\vphi_*+\delta\vphi\rightarrow\vphi,\;\dot\delta\vphi\rightarrow\dot\vphi,\;\ddot\delta\vphi\rightarrow\ddot\vphi,$$ where the perturbation $\delta\vphi\ll 1$, i. e., $\ddot\delta\vphi\sim\dot\delta\vphi\sim\delta\vphi\sim O(1)$, etc. Up to linear terms, the corresponding  KG equation for the perturbation reads: $$\ddot\delta\vphi+3H\dot\delta\vphi=-\frac{(\lambda+1)\left[2M^2(\lambda-1)\right]^\frac{1}{\lambda+1}}{3+2\omega_\textsc{bd}}\,\rho_m^\frac{\lambda}{\lambda+1}\,\delta\vphi.$$ 

The problematic terms in the above equation are the time--dependent terms $H(t)$ and $\rho_m(t)$. Since the RHS term can be written as $-m^2_{\vphi_*}\delta\vphi$, where $$m^2_{\vphi_*}=\frac{(\lambda+1)\left[2M^2(\lambda-1)\right]^\frac{1}{\lambda+1}}{3+2\omega_\textsc{bd}}\,\rho_m^\frac{\lambda}{\lambda+1},$$ is the effective mass squared of the perturbation, then, as long as $\rho_m$ is a function of the cosmic time, the effective mass is so: $m_{\vphi_*}=m_{\vphi_*}(t)$. But then, this mass would not be a standard mass of an elementary perturbation in SF--BD theory, since any physically meaningful mass parameter in the SF/JF formulations of the Brans--Dicke theory should be independent of the spacetime point. Otherwise, as mentioned, coexistence of particles with constant mass and elementary perturbations with point--dependent mass, would entail serious problems with the equivalence principle.

As a consequence of the above analysis, in a cosmological setting, the chameleon effect makes sense only during small enough intervals of cosmic time, so that the background energy density may be considered as a constant. Only if the background were the quantum vacuum with constant density $\rho_\text{vac}$ during the course of the cosmic evolution, the chameleon effect may have cosmological sense. During a small time--interval $\delta t\sim O(1)$, $H(t)\approx H_0(1+\alpha_0\delta t)$, and $\rho_m(t)\approx\rho_m^0(1+\beta_0\delta t)$. Substituting these equations back into the SF--KG equation above, and keeping the terms linear in the perturbations $\delta\vphi\sim\delta t\sim O(1)$, one obtains:

\bea &&\ddot\delta\vphi+3H_0\dot\delta\vphi\approx-\left(m^0_{\vphi_*}\right)^2\,\delta\vphi,\label{damped-osc}\\
&&\left(m^0_{\vphi_*}\right)^2=\frac{(\lambda+1)\left[2M^2(\lambda-1)\right]^\frac{1}{\lambda+1}}{3+2\omega_\textsc{bd}}\,\left(\rho^0_m\right)^\frac{\lambda}{\lambda+1}.\nonumber\eea 

Thanks to the friction term $3H_0\,\dot\delta\vphi$, the solution of the linear approximation to the SF--KG equation above, are the damped oscillations:

\bea \delta\vphi(t)=C\,e^{-\frac{3H_0}{2}\,t}\,\sin\left(\omega\,t+\Phi\right),\label{damped-osc-sol}\eea where $\Phi$ is an arbitrary phase which depends on the initial conditions chosen, $C$ is an integration constant, and $$\omega=\sqrt{\left(m^0_{\vphi_*}\right)^2-\left(\frac{3H_0}{2}\right)^2}.$$ 

This solution is valid only during small enough interval of the cosmic time $\delta t\ll 2/3H_0\sim$ lifetime of the universe. Hence, the damped oscillations that transmute the SF--BD theory into Einstein's GR with a cosmological constant, which are valid for any chameleon potential with a minimum, can not be an equilibrium configuration in the equivalent phase space. These can be, at most, a local attractor or, even, a saddle point. A global attractor in the phase space is a stable equilibrium point which does not depend on how far into the past the initial conditions for an specific orbit are given. As shown in section \ref{dyn-syst-sec}, this is obtained only for the exponential potential $U(\vphi)\propto\exp\vphi$, or for potentials that asymptote to $\exp\vphi$.

Our conclusion is that one should trust the results of the study based on the dynamical systems theory (see section \ref{dyn-syst-sec} and the references \cite{hrycyna, bd-quiros}), where the chameleon effect is not explicitly considered, over the results based on the above explained transmutation, since in a cosmological setting the latter effect is valid only during short stages of the cosmic evolution, where the background density of matter may be considered almost a constant value and, consequently, it can not give rise to a global attractor. This does not mean that the chameleon effect is not physically meaningful, but rather that, in a cosmological setting, its consequences are of limited reach. 

Our results here differ from those in Ref. \cite{cosmo-cham}, where the cosmological chameleon is also investigated, but not within the context of the BD theory, and not in the Jordan/string frames, but in the Einstein frame. In that reference, the chameleon obviously violates the Einstein's equivalence principle (EEP), since its effective mass evolves during the course of the cosmic evolution. This might be related with the fact that in \cite{cosmo-cham}  (see aslo \cite{cham}), the physically meaningful matter density is not the one measured by co--moving observers in the EF (this is not the conserved one in the EF), neither the conformal one, but a density which does not depend on the dilaton. In \cite{cosmo-cham}, the matter fields couple to the conformal metric $g^{(i)}_{\mu\nu}=\exp(2\beta_i\phi/M_\textsc{pl})\,g_{\mu\nu}$, while the density of the non--relativistic fluid measured by EF co--moving observers is denoted by $\tilde\rho_i$. It is assumed that what matters is the $\phi$--independent density $\rho_i=\tilde\rho_i\exp(3\beta_i\phi/M_\textsc{pl})$, which is the one conserved in the EF. While this choice may not be unique, in the Jordan as well as in the string frames of the Brans--Dicke theory one does not have this ambiguity: the matter density measured by JF(SF) co--moving observers $\rho$, is the one conserved in the Jordan/string frames and, additionally, does not depend on the BD--field (on the dilaton).


\section{The cosmological constant problem}\label{ccp-sec}

The chameleon transmutation of BD theory into GR within regions of constant density of matter, poses a new problem that can be stated as follows. Let us imagine two separate spatial regions: one region with mean (constant) density of matter $\rho_0$, and another region with (also constant) density $\rho_\infty$. Let the chameleon potential $U_\text{ch}(\vphi)$ to have different minimums at each one of these regions: $\vphi_0$ and $\vphi_\infty$, respectively. We have that $$e^{\vphi_0}=\left[\frac{\rho_0}{2M^2(\lambda-1)}\right]^\frac{1}{\lambda+1},\;e^{\vphi_\infty}=\left[\frac{\rho_\infty}{2M^2(\lambda-1)}\right]^\frac{1}{\lambda+1}.$$ 

The resulting problem is that we will have two different values of the measured ``Planck mass'' $M^2_\textsc{pl}=\exp\vphi$. Their ratio can be written as: 

\bea \frac{\left(M^0\right)^2_\textsc{pl}}{\left(M^\infty\right)^2_\textsc{pl}}=\left(\frac{\rho_0}{\rho_\infty}\right)^\frac{1}{\lambda+1}.\label{mpl-ratio}\eea 

An also related problem is the well--known (old) cosmological constant problem \cite{c-c-ratra-peebles, c-c-weinberg, c-c-padmanabhan, c-c-zlatev, c-c-carroll}: why the present energy density of vacuum: $\rho^\text{cosm}_\text{vac}\sim 10^{-48}$ GeV$^4$, is so small compared with the expected (theoretical) value: $\rho^\textsc{pl}_\text{vac}\sim 10^{72}$ GeV$^4$? In our example the present (cosmological) energy density will be identified with the mean density $\rho_\infty$, meanwhile, the theoretical expected value may be the mean density $\rho_0$. In consequence, we have that 

\bea \frac{\Lambda_0}{\Lambda_\infty}=\frac{U(\vphi_0)}{U(\vphi_\infty)}=\left(\frac{\rho_0}{\rho_\infty}\right)^\frac{\lambda}{\lambda+1},\label{c-c-p}\eea where $\Lambda_0=U(\vphi_0)$ and $\Lambda_\infty=U(\vphi_\infty)$, are the values of the cosmological constant measured in each one of the regions. 

Hence, we have two related problems that, for the exponential potential case, are expressed by equations (\ref{mpl-ratio}) and (\ref{c-c-p}), respectively, which are associated with the occurrence of the chameleon effect in the SF--BD theory. A way out may be to assume large $\lambda\gg 1$ ($\lambda\gg 119$ to be precise). In this case we have 

\bea \frac{\left(M^0\right)^2_\textsc{pl}}{\left(M^\infty\right)^2_\textsc{pl}}=10^\frac{120}{\lambda+1}\approx 1,\label{mpl-ratio-l}\eea i. e., the measured value of the Planck mass is almost the same in both regions. At the same time, there can be a huge difference between the value of the cosmological constant measured in one region when compared with the value measured in the other region:

\bea \frac{\Lambda_0}{\Lambda_\infty}=10^\frac{120\lambda}{\lambda+1}\approx 10^{120},\label{c-c-p-l}\eea as observed. In this case the origin of the cosmological constant problem may be attributed to the chameleon effect itself. 

Notice that, in order to get a consistent picture, here we have to set $\lambda\gg 1$, while the global GR--de Sitter attractor occurs for $\lambda=1$ (see section \ref{dyn-syst-sec}).


\section{jordan/string and einstein frames}\label{frames-sec}

The conformal transformations controversy \cite{c-t-dicke, c-t-faraoni, c-t-faraoni-nadeau} is still a subject of debate \cite{c-t-polacos, c-t-geometry, c-t-debate}. Under a conformal transformation of the metric:

\bea \bar g_{\mu\nu}=\Omega^2g_{\mu\nu}\;\Rightarrow\;\sqrt{|g|}=\Omega^{-4}\sqrt{|\bar g|},\label{conf-transf-m}\eea with $\Omega^2=\phi$, the JF--DB theory, depicted by the action (\ref{bd-action}), is transformed into the Einstein frame formulation of BD theory:

\bea &&S^\phi_\textsc{ef}=\int d^4x\sqrt{|\bar g|}\left[\bar R-\left(\omega_\textsc{bd}+\frac{3}{2}\right)\left(\frac{\der\phi}{\phi}\right)^2\right.\nonumber\\
&&\left.\;\;\;\;\;\;\;\;\;\;\;\;-2\phi^{-2}V(\phi)\right]+2\int d^4x\sqrt{|\bar g|}\,\phi^{-2}{\cal L}_m,\label{e-frame-bd-action}\eea where, as before, ${\cal L}_m={\cal L}_m\left(\chi,\bar\der\chi,\phi^{-1}\bar g_{\mu\nu}\right)$, is the Lagrangian of the matter fields $\chi$, which are coupled to the JF metric $g_{\mu\nu}=\phi^{-1}\bar g_{\mu\nu}$. 

There is an ongoing debate on whether theory (\ref{bd-action}) or theory (\ref{e-frame-bd-action}), has the physical meaning \cite{c-t-dicke, c-t-faraoni, c-t-faraoni-nadeau, c-t-polacos, c-t-geometry, c-t-debate}. In this regard, the chameleon effect has been investigated, exclusively, in the Einstein frame version of the scalar-tensor theory. This fact leaves room for a certain suspicion on the real physical meaning of this effect. This is why we are discussing in the present paper the chameleon effect from the perspective of the Jordan and string frames formulation of Brans-Dicke theory. Although we shall not discuss here the very warped subject of the conformal transformation's issue, nevertheless, certain features related with these transformations will be discussed within the context of the chameleon effect.

\subsection{String--frame BD theory}

The string frame formulation of the dilaton--gravity action \cite{wands-rev} admits, in principle, that different matter fields may couple to different conformal metrics, instead of the SF metric $g_{\mu\nu}$: $${\cal L}_m\left(\chi_{(i)},\der\chi_{(i)},g^{(i)}_{\mu\nu}\right),$$ where the matter fields $\chi_{(i)}$ couple to the conformal metrics: 

\bea g^{(i)}_{\mu\nu}=e^{\beta_i\vphi}g_{\mu\nu}.\label{conf-i-metric}\eea The $\beta_i$-s are different coupling strengths of order unity. 

Let us assume, for simplicity, a single matter species, so that $g_{\mu\nu}^{(1)}=e^{\beta\vphi}g_{\mu\nu}$. The SF dilaton--gravity action plus the matter fields reads:

\bea &&S^\vphi_\textsc{sf}=\int d^4x\sqrt{|g|}\,e^\vphi\left[R-\omega_\textsc{bd}\left(\der\vphi\right)^2-2U(\vphi)\right]\nonumber\\
&&\;\;\;\;\;\;\;\;\;\;+2\int d^4x\sqrt{|g|}\,e^{2\beta\vphi}{\cal L}_m\left(\chi,\der\chi,e^{\beta\vphi}g_{\mu\nu}\right),\label{s-frame-dilaton-action}\eea where we have dropped the subscript $i$, so that, as before, $\chi$ stands collectively for the matter fields. For $\beta=0$ we recover the string frame formulation of the BD theory given by the action (\ref{dilaton-action}).

Under a conformal transformation of the metric (\ref{conf-transf-m}) with $\Omega^2=e^\vphi$, the above action is transformed into the EF action:

\bea &&S^\vphi_\textsc{ef}=\int d^4x\sqrt{|\bar g|}\left[\bar R-\left(\omega_\textsc{bd}+\frac{3}{2}\right)\left(\bar\der\vphi\right)^2-2\bar U(\vphi)\right]\nonumber\\
&&\;\;+2\int d^4x\sqrt{|\bar g|}\,e^{2(\beta-1)\vphi}{\cal L}_m\left(\chi,\bar\der\chi,e^{(\beta-1)\vphi}\bar g_{\mu\nu}\right),\label{e-frame-dilaton-action}\eea where $\bar U(\vphi)=e^{-\vphi}U(\vphi)$. Besides, under (\ref{conf-transf-m}) ($\Omega^2=e^\vphi$), the stress-energy tensor of the matter fields $$\bar T^{(m)}_{\mu\nu}=-2\frac{\der\left[\sqrt{|\bar g|}\,e^{2(\beta-1)\vphi}{\cal L}_m\left(\chi,\bar\der\chi,e^{(\beta-1)\vphi}\bar g_{\mu\nu}\right)\right]}{\sqrt{|\bar g|}\der g^{\mu\nu}},$$ transforms like $\bar T^{(m)}_{\mu\nu}=e^{-\vphi}T^{(m)}_{\mu\nu}$, where $$T^{(m)}_{\mu\nu}=-2\frac{\der\left[\sqrt{|g|}\,{\cal L}_m\left(\chi,\der\chi,g_{\mu\nu}\right)\right]}{\sqrt{|g|}\der g^{\mu\nu}}.$$ 

The coupling of the matter fields to the conformal metric and not to the SF metric $g_{\mu\nu}$, means that the stress--energy is not separately conserved in the string frame, but that there is exchange of energy--momentum between the matter fields $\chi$ and the dilaton: $$\nabla^\mu T^{(m)}_{\mu\nu}=\frac{\beta}{2}\,\der_\nu\vphi\,T^{(m)},$$ where $T^{(m)}=g^{\mu\nu}T^{(m)}_{\mu\nu}$ is the trace of the stress-energy tensor of matter. Under the conformal transformation (\ref{conf-transf-m}) with $\Omega^2=e^\vphi$, the above ``continuity'' equation is transformed into: $$\bar\nabla^\mu\bar T^{(m)}_{\mu\nu}=\frac{\beta-1}{2}\,\der_\nu\vphi\bar T^{(m)},$$ so that the energy is not conserved in the Einstein frame neither.

Setting $\beta=0$ in the former equations, amounts to the assumption that the matter fields are minimally coupled to the SF metric, which corresponds to the standard Brans--Dicke theory depicted by the dilatonic (SF) variables. This means, in turn, that, in the string frame, the standard conservation equation is satisfied: $\nabla^\mu T^{(m)}_{\mu\nu}=0$, while in the EF the stress--energy is not separately conserved:

\bea \bar\nabla^\mu\bar T^{(m)}_{\mu\nu}=-\frac{1}{2}\der_\nu\vphi\bar T^{(m)}.\label{ef-dil-cons-eq}\eea Instead $\bar\nabla^\mu\bar T^{(m)}_{\mu\nu}+\bar\nabla^\mu\bar T^{(\vphi)}_{\mu\nu}=0,$ where $$\bar T^{(\vphi)}_{\mu\nu}=\left(\omega_\textsc{bd}+\frac{3}{2}\right)\left[\der_\mu\vphi\der_\nu\vphi-\frac{1}{2}\bar g_{\mu\nu}\left(\bar\der\vphi\right)^2\right]-\bar g_{\mu\nu}\bar U(\vphi),$$ so that $$\bar\nabla^\mu\bar T^{(\vphi)}_{\mu\nu}=\left[\left(\omega_\textsc{bd}+\frac{3}{2}\right)\bar\nabla^2\vphi-\der_\vphi\bar U\right]\der_\nu\vphi.$$ These equations lead to the EF--KG equation:

\bea \bar\nabla^2\vphi=\frac{2}{3+2\omega_\textsc{bd}}\left[\der_\vphi\bar U+\frac{1}{2}\,\bar T^{(m)}\right],\label{ef-dilaton-kg-eq}\eea which can be obtained, alternatively, by performing the conformal transformation (\ref{conf-transf-m}), in Eq. (\ref{dilaton-kg}), with $\Omega^2=e^\vphi$.

If choose $\beta=1$ in the equations above, then the stress--energy of the matter fields is conserved in the Einstein frame, and the EF action (\ref{e-frame-dilaton-action}) coincides with the standard Einstein-Hilbert action of general relativity. In this case the action (\ref{s-frame-dilaton-action}) will amount to the string frame (conformal) formulation of general relativity \cite{quiros-prd-2000}.


\section{effective mass of the chameleonic dilaton in the Einstein frame}\label{ef-sec}

Let us to compute the effective mass  $\bar m^2_\text{eff}$, of the chameleonic dilaton in the Einstein frame. This is the mass which determines the range of the Yukawa--like modification of the Newtonian potential $$\Delta\bar U_N\propto-\frac{e^{-r/\bar\lambda_\text{eff}}}{r},$$ where $\bar\lambda_\text{eff}=\bar m^{-1}_\text{eff}$ is the Compton length of the dilaton in the Einstein frame. 

For definiteness we will consider the Brans-Dicke theory, which, in the string frame is given by (\ref{dilaton-action}), i. e., by the action (\ref{s-frame-dilaton-action}) with $\beta=0$. Hence, in the EF action (\ref{s-frame-dilaton-action}), and in the related equations, one has to set $\beta=0$ as well. This means that we will be dealing with the EF--KG equation (\ref{ef-dilaton-kg-eq}). Following the same procedure of sections \ref{bd-mass-sec}, \ref{cham-mass-sec}, and \ref{dilaton-mass-sec}, we get that the effective mass squared of the dilaton in the Einstein frame is given by: $\bar m^2_\text{eff}=\der^2_\vphi\bar U_\text{ch}$. 

In the Einstein frame the chameleon potential is depicted by:

\bea \bar U_\text{ch}=\frac{2}{3+2\omega_\textsc{bd}}\left[\bar U(\vphi)-\frac{1}{4}\,\bar T^{(m)}\right],\label{ef-cham-pot}\eea where there is subtlety that is to be explained. Assuming that in the Jordan/string frames the masses of particles $m_0$, and the related quantities such as the density of matter $\rho_0$, do not depend on spacetime position, their conformal cousins in the Einstein frame $\bar m_0(\vphi)=\Omega^{-1}m_0=\exp(-\vphi/2)m_0$, and $\bar\rho_0(\vphi)=\Omega^{-4}\rho_0=\exp(-2\vphi)\rho_0$, will be point--dependent quantities. Hence, for instance,

\bea \der_\vphi\bar\rho_0=-2\bar\rho_0,\;\der^2_\vphi\bar\rho_0=4\bar\rho_0,\;...,\label{der-rho}\eea etc. Then, since in the EF what the co--moving observers with 4-velocity $\bar u^\mu=\delta^\mu_0$ measure, is the energy density $$\bar\rho_0=\bar T^{(m)}_{\mu\nu}\bar u^\mu\bar u^\nu=\bar T^{(m)}_{00},$$ and assuming, as before, that in the EF we have a pressureless (non--relativistic) fluid $\bar T^{(m)}=-\bar\rho_0$, the derivative of the chameleon potential (\ref{ef-cham-pot}): $$\der_\vphi\bar U_\text{ch}=\frac{2}{3+2\omega_\textsc{bd}}\left[\der_\vphi\bar U(\vphi)-\frac{1}{2}\,\bar\rho_0(\vphi)\right],$$ coincides with the RHS of Eq. (\ref{ef-dilaton-kg-eq}). 

Let us consider the SF exponential potential (\ref{exp-pot}), so that, in the Einstein frame:

\bea \bar U(\vphi)=M^2\,e^{(\lambda-1)\vphi}.\label{ef-exp-pot}\eea The resulting EF chameleonic potential (\ref{ef-cham-pot}), reads:

\bea \bar U_\text{ch}=\frac{2M^2}{3+2\omega_\textsc{bd}}\left[e^{(\lambda-1)\vphi}+\frac{\bar\rho_0(\vphi)}{4M^2}\right].\label{ef-cham-exp-pot}\eea It is a minimum at the value $\vphi_*$ which solves the equation $$e^{(\lambda-1)\vphi_*}=\frac{\bar\rho_0(\vphi_*)}{2M^2(\lambda-1)},$$ i. e., since $\bar\rho_0(\vphi)=\exp(-2\vphi)\rho_0$, where $\rho_0$ is a constant, then $$\vphi_*=\frac{1}{\lambda+1}\ln\left[\frac{\rho_0}{2M^2(\lambda-1)}\right].$$ This position of the minimum in the $\vphi$--direction coincides with the one in the string frame in Eq. (\ref{sf-min}). We have that: 

\bea \bar\rho_0(\vphi_*)=\left[4M^4(\lambda-1)^2\right]^\frac{1}{\lambda+1}\,\rho_0^\frac{\lambda-1}{\lambda+1}.\label{rho-0}\eea 

The effective chameleonic mass squared of the dilaton in the Einstein frame $\bar m^2_\text{eff}=\bar m^2_{\bar\vphi_*}$, is given by the following expression:

\bea \bar m^2_{\vphi_*}=\der^2_\vphi\bar U_\text{ch}(\vphi_*)=\frac{(\lambda+1)\,\bar\rho_0(\vphi_*)}{3+2\omega_\textsc{bd}},\label{ef-cham-dil-mass}\eea where we prefer to keep the dependence on $\bar\rho_0$, which is the density of matter measured by EF co--moving observers, instead of on the SF constant $\rho_0$. In other words, the effective EF mass squared of the dilaton depends linearly with the density of the surrounding matter $\bar m^2_{\vphi_*}\propto\bar\rho_0$, as measured by EF co--moving observers.

If desired, the EF mass squared $\bar m^2_{\vphi_*}$ in Eq. (\ref{ef-cham-dil-mass}) can be straightforwardly written in terms of the SF constant $\rho_0$ as well, by substituting (\ref{rho-0}) in Eq. (\ref{ef-cham-dil-mass}): 

\bea \bar m^2_{\vphi_*}=\frac{(\lambda+1)\left[4M^4(\lambda-1)^2\right]^\frac{1}{\lambda+1}}{3+2\omega_\textsc{bd}}\,\rho_0^\frac{\lambda-1}{\lambda+1}.\label{ef-mass-rho}\eea 

This equation confirms the transformation law for the mass under (\ref{conf-transf-m}): $\bar m_{\vphi_*}=\Omega^{-1}m_{\vphi_*}$. Actually, under the conformal transformation (\ref{conf-transf-m}) with $\Omega^2=\exp\vphi$, $$\bar m^2_{\vphi_*}=e^{-\vphi_*}m^2_{\vphi_*}=\left[\frac{\rho_0}{2M^2(\lambda-1)}\right]^{-\frac{1}{\lambda+1}}m^2_{\vphi_*},$$ which, if substitute $m^2_{\vphi_*}$ from Eq. (\ref{exp-cham-mass}), yields (\ref{ef-mass-rho}).


\section{Conformal frames and the chameleon effect}\label{conf-frames-sec}

The controversy on which of the conformal frames where the Brans--Dicke theory can be formulated: the JF/SF or the EF, is physical, started with the paper by Dicke \cite{c-t-dicke}, who stated that these are both equivalent representations of the same theory. However, no matter how trivial the question seems, the debate has not ceased for over the last 53 years \cite{c-t-faraoni, c-t-faraoni-nadeau, c-t-polacos, c-t-geometry, c-t-debate}. In the present section we want to extend the debate to the discussion on the physical implications of the chameleon effect for the BD theory.

\subsection{The SF and the EF are equivalent representations of the same theory}

According to one of the most popular points of view on the meaning of the conformal transformations of the metric \cite{c-t-dicke, c-t-faraoni-nadeau}, both conformal frames: the SF/JF and the EF, are physically equivalent. The basic idea is that the two conformal frames are physically equivalent provided that in the EF the units of time, length, mass, and derived quantities are allowed to scale with appropriate powers of the conformal factor $\Omega$. If one adheres to this point of view, as we temporarily do here, one must accept that, instead of a system of units rigidly attached to the spacetime, the Einstein frame contains a system of units that depend on the spacetime point \cite{c-t-faraoni-nadeau} (for an alternative point of view on the equivalence see \cite{c-t-geometry}).

Let us consider that the mass of a given field, say, $m_f$, is a constant in the string frame. Under a conformal transformation $\bar g_{\mu\nu}=\Omega^2 g_{\mu\nu}$ $\Rightarrow\;\bar m_f=\Omega^{-1}m_f$, so that, in the EF the mass of the field will be point--dependent, provided that in the SF the mass of the particle $m_f$ is a constant. The fine point is that, in an experiment what one measures is the ratio $\bar m_f/\bar m_u$ between the mass of the field and an arbitrarily chosen mass unit $\bar m_u$. Hence, in the EF it is the conformal invariant ratio 

\bea \frac{\bar m_f}{\bar m_u}=\frac{\Omega^{-1}m_f}{\Omega^{-1}m_u}=\frac{m_f}{m_u},\label{ratio}\eea what really matters \cite{c-t-faraoni-nadeau}. As a consequence, a measurement of the field's mass yields a same value in the EF and in the SF. In other words, according to the present point of view, the given measurement can not differentiate between the different conformal frames. In order to illustrate this case, we take, for instance, the exponential potential $U(\vphi)\propto\exp(\lambda\vphi)$, and use the equation for the EF effective mass squared (\ref{ef-mass-rho}), instead of (\ref{ef-cham-dil-mass}). In this case one has,

\bea &&\frac{\bar m_{\vphi_*}}{\bar m_u}=\sqrt\frac{\lambda+1}{3+2\omega_\textsc{bd}}\frac{\left[2M^2(\lambda-1)\right]^\frac{1}{\lambda+1}\rho_0^\frac{\lambda-1}{2(\lambda+1)}}{\Omega^{-1}(\vphi_*)\,m_u}\nonumber\\
&&=\sqrt\frac{\lambda+1}{3+2\omega_\textsc{bd}}\frac{\left[\sqrt{2M^2(\lambda-1)}\right]^\frac{1}{\lambda+1}\rho_0^\frac{\lambda}{2(\lambda+1)}}{m_u}=\frac{m_{\vphi_*}}{m_u}.\nonumber\eea 

In this case the chameleon effect is the same independent of the frame, and, provided that the observational bounds are met in one frame within a certain accuracy, these will be met in its conformal partner with the same accuracy as well. The contrary statement is also true.

\subsection{The SF and the EF nest different theories}

There is a radically different point of view on the (in)equivalence of the different conformal formulations of BD theory. According to this viewpoint, the SF--BD theory and the EF--BD theory are totally different theories: i) same metric affinity (pseudo--Riemann spaces), ii) different equations of motion, i. e., different laws of gravity, iii) different experimental bounds to meet. In this case the conformal transformations of the metric are understood just as the deformation of one theory, that leads to the conformal one. 

In what regards to the chameleon effect within the BD theory, this viewpoint is supported by the equation (\ref{ef-cham-dil-mass}) for the EF dilaton's effective mass squared. It is understood that, what really matters in the Einstein frame, are the quantities that EF co--moving observers with 4--velocity $\bar u^\mu=\delta^\mu_0$, measure. Co--moving observers in the EF see, for instance, exchange of energy--momentum between the dilaton fluid and the standard matter fields, i. e., they ``see'' a fifth force which makes the motion of point--like particles to depart from geodesic motion in the EF metric $\bar g_{\mu\nu}$. This is to be contrasted with what SF co--moving observers with 4--velocity $u^\mu=\delta^\mu_0$, see: the stress--energy tensor of each one of the matter species, is separately conserved. In other words, the SF co--moving observers do not see any fifth force effect. Besides, while the EF co--moving observers measure the energy density $\bar\rho_0=\bar T^{(m)}_{00}=\bar T^{(m)}_{\mu\nu}\bar u^\mu\bar u^\nu$, the JF co--moving observers measure the constant energy density: $\rho_0=T^{(m)}_{00}$.

If assume that the present viewpoint is correct, i. e., that $\bar m_{\vphi_*}\propto\sqrt{\bar\rho_0}$, then the EF--BD theory with exponential potential $\bar U(\vphi)\propto\exp[(\lambda-1)\vphi]$, which is the EF cousin of $U(\vphi)\propto\exp(\lambda\vphi)$, improves the estimates of section \ref{estimate-sec} by several orders of magnitude. Actually, working as we did in section \ref{estimate-sec}, we obtain that

\bea \frac{\bar m^\text{cosm}_{\vphi_*}}{\bar m^\text{atm}_{\vphi_*}}=\sqrt\frac{\bar\rho_0^\text{crit}}{\bar\rho_0^\text{atm}}\sim 10^{-14},\label{ef-estimate}\eea which is to be contrasted wit (\ref{estimate}). In the above equation we have considered that $\bar\rho_0^\text{crit}\sim 10^{-31}$ g/cm$^3$, and $\bar\rho_0^\text{atm}\sim 10^{-3}$ g/cm$^3$, are the critical energy density of the universe and the mean density of the Earth atmosphere, respectively, which are measured by EF co--moving observers. 

In spite of the very discrete improvement of the EF--based estimate over the SF--based one for the exponential potential, it might represent a convenient observational signature, that may differentiate between the SF and EF formulations of the Brans--Dicke theory.

\subsection{The effective field--theoretical mass}\label{eff-ft-mass-subsec}

According to Eq. (\ref{eff-dilaton-mass}), the effective field--theoretical mass squared of the dilaton in the string frame is given by $$m^2_\vphi(\vphi)=\frac{2}{3+2\omega_\textsc{bd}}\left[\der^2_\vphi U-\der_\vphi U-\frac{T^{(m)}}{2}\,e^{-\vphi}\right],$$ while, in the Einstein frame, given the chameleon potential (\ref{ef-cham-pot}), it is given by $$\bar m^2_\vphi(\vphi)=\der^2_\vphi\bar U_\text{ch}=\frac{2}{3+2\omega_\textsc{bd}}\left[\der^2_\vphi\bar U-\bar T^{(m)}\right],$$ where it has been taken into account that $\der_\vphi\bar T^{(m)}=-2\bar T^{(m)}$. There is no way in which one of the above equations can be obtained from the other one by means of a conformal transformation of the metric. Actually, if in the first equation above make the substitutions: $$U(\vphi)=\Omega^2\bar U(\vphi),\;T^{(m)}=\Omega^4\bar T^{(m)},\;\Omega^2=e^\vphi,$$ one obtains that 

\bea m^2_\vphi(\vphi)=\frac{2\,e^\vphi}{3+2\omega_\textsc{bd}}\left[\der^2_\vphi\bar U+\der_\vphi\bar U-\frac{\bar T^{(m)}}{2}\right],\label{eq}\eea that can not be written in the form of the corresponding conformal transformation of a mass parameter: $m^2_\vphi=\Omega^2\bar m^2_\vphi=e^\vphi\,\bar m^2_\vphi$. Only if the chameleon potential (in either frame) is a minimum at some $\vphi=\vphi_*$, the obtained effective mass retains the sense of an ordinary mass parameter. As a matter of fact, if $\bar U_\text{ch}$ in (\ref{ef-cham-pot}) has a minimum, this means that, at the minimum $$\der_\vphi\bar U_\text{ch}=0\;\Rightarrow\;\der_\vphi\bar U=-\frac{1}{2}\,\bar T^{(m)},$$ so that, the equation (\ref{eq}) can be written as: $$m^2_\vphi=\frac{2\,e^\vphi}{3+2\omega_\textsc{bd}}\left[\der^2_\vphi\bar U-\bar T^{(m)}\right]=e^\vphi\,\bar m^2_\vphi.$$ 

This means that the physical meaning is with the effective mass which is obtained from second derivatives of a chameleon potential evaluated at the minimum. In contrast, the effective field--theoretical mass has limited physical implications.


\section{discussion and conclusion}\label{discu-conclu-sec}

According to Eq. (\ref{exp-cham-pot}), for the exponential potential $U(\vphi)\propto\exp(\lambda\vphi)$, the effective mass squared of the dilaton is given by: 

\bea m^2_{\vphi_*}=\frac{(\lambda+1)\left[2(\lambda-1)M^2\right]^\frac{1}{\lambda+1}}{3+2\omega_\textsc{bd}}\,\rho^\frac{\lambda}{\lambda+1}.\label{mass-lambda}\eea The dependence of the dilaton's effective mass on the density of the environment $m_{\vphi_*}=m_{\vphi_*}(\rho)$, is the basis for the chameleon effect. The interest in the latter as an effective screening mechanism, resides in that the dilaton field might have impact in the cosmological dynamics and, yet, it might be effectively screened from the solar system and terrestrial experiments. This is in contrast to the mass due to an effective potential which does not depend on $\rho$. In this case, provided that the effective mass is small enough, the dilaton may modify the cosmological dynamics. However, since the mass is small no matter where the experiments with the dilaton are performed, one has to care about the stringent bounds coming from the terrestrial fifth--force experiments, the solar system experiments to test the equivalence principle, etc. 

In the case of the Brans--Dicke theory, one interesting application of the chameleon effect is related with the possible relaxation of the stringent lower bounds on the BD coupling parameter $\omega_\textsc{bd}>40000$. In order to illustrate the way such a relaxation may occur, let us assume, as we did in section \ref{estimate-sec}, that the dilaton is immersed in the earth atmosphere with mean density $\rho^\text{atm}\sim 10^{-3}$ g/cm$^3$. Assume, besides, that the 1 mm lower bound on the modification of the Newton's law of gravitation applies, i. e., that $m_{\vphi_*}^{-1}\sim 1$ mm, and, for sake of definiteness, let us set $\lambda=3$. This value of the free parameter $\lambda$, corresponds to the quartic potential $V(\phi)\propto\phi^4$ in terms of the Jordan frame variables. If substitute the above numbers into Eq. (\ref{mass-lambda}) 

\bea &&m^2_{\vphi_*}=\frac{4\left(4M^2\right)^{1/4}\rho^{3/4}}{3+2\omega_\textsc{bd}}\;\Rightarrow\nonumber\\
&&m^{-1}_{\vphi_*}[\text{mm}]\approx\frac{0.1\sqrt{3+2\omega_\textsc{bd}}}{2(4M^2)^{1/8}\left(\rho[\text{gr/cm}^3]\right)^{3/8}},\nonumber\eea one gets that 

\bea \omega_\textsc{bd}\approx 1.6\sqrt{M}-1.5,\label{omega-estimate}\eea meaning that for $M\sim 1$, one can have that the BD theory with an $\omega_\textsc{bd}$ of order unity may be the correct description of gravity, without conflict with the existing experimental bounds, coming from terrestrial and solar system tests of the equivalence principle, fifth force and modifications of the Newtonian gravitational potential.

The interesting question is whether the same theory can explain the observational data of cosmological origin as well. Worth noticing that the effective mass squared $m^2_{\vphi_*}$ in Eq. (\ref{mass-lambda}), vanishes if $\lambda=1$, i. e., for the specific exponential $U(\vphi)\propto\exp\vphi$. This result is independent of the conformal frame: in the EF we have that (see Eq. (\ref{ef-mass-rho})), $$\bar m^2_{\vphi_*}=\frac{(\lambda+1)\left[4M^4(\lambda-1)^2\right]^\frac{1}{\lambda+1}}{3+2\omega_\textsc{bd}}\,\rho^\frac{\lambda-1}{\lambda+1},$$ so that, the EF effective mass squared of the dilaton vanishes if $\lambda=1$ as well. As a consequence, for $\lambda=1$, the chameleon effect does not arise in neither frame. This means that the dilaton--mediated interaction of matter is of long--range (it is not screened), and we would have had detected the BD dilaton in terrestrial and solar system experiments. If, on the contrary, one assumes that $\lambda\neq 1$, as we have shown above, the chameleon effect arises but, then the GR--de Sitter solution is not an equilibrium point of the dynamical system (\ref{x-xi-ode-vac}), neither of the simpler ODE (\ref{x-ode-vac-exp}). This means, in turn, that the $\Lambda$CDM model is not an attractor of the vacuum Brans--Dicke cosmology.\footnote{As shown in the Ref. \cite{bd-quiros}, this result is also true for Brans--Dicke cosmology with pressureless matter.} The absence of the GR--de Sitter critical point entails that, in case it were an exact solution of the BD cosmological field equations (\ref{efe}), it were a non--generic unstable one, without interest for describing a long--lasting period of the cosmological evolution. 

Another essential aspect of the chameleon mechanism is related with the fact that a convenient effective chameleon potential $U_\text{ch}(\vphi,\rho)$, is a minimum at some $\vphi_*$, where $$\der_\vphi U_\text{ch}(\vphi_*)=0,\;m^2_{\vphi_*}=\der^2_\vphi U_\text{ch}(\vphi_*)>0.$$ In a cosmological setting this entails that the dilaton does damped oscillations around the expectation value $\vphi_*$, until it stabilizes at the minimum of the potential. As a consequence, the Brans--Dicke theory with a chameleonic dilaton renders general relativity at the minimum of the potential. Notice, however, that, in general, the density of the environment $\rho=\rho(t)$ is a function of the cosmic time. This is particularly true if one considers non--static distributions of matter. In this more general case what one has is not a point of minimum, but a whole curve $\vphi_*(t)=\vphi_*(\rho(t))$. Hence, the dilaton does not acquire an expectation value, but an expectation curve, whatever this means. In such a case, $\dot\vphi_*(t)\neq 0$ and one does not end up with general relativity but with Brans--Dicke theory instead. Hence, a necessary condition for the GR--de Sitter space to be a local attractor (or a saddle point) of the BD cosmology is that the density of matter $\rho$ be a constant within a given region. 

One gets the following big picture: Provided that the laws of cosmological expansion are governed by the Brans--Dicke theory with the potential $U(\vphi)$, if there where regions ${\cal R}_i$ in our universe where the density of matter $\rho_i$ were a constant (perhaps a different constant in each region), and assuming that within each one of such regions the potential $U(\vphi)$ were a minimum at some $\vphi_{i*}=\vphi_{i*}(\rho_i)$ (different for each region also), then, thanks to the chameleon effect, after different periods of damped oscillations around the different minimums $\vphi_{i*}$, the BD theory transmutes into GR--de Sitter gravity within each one of these regions, with different sets of fundamental constants $(M^2_{\textsc{pl},i},\Lambda_i)$. Yet, if the potential $U(\vphi)$ were one that asymptotes to the exponential $U(\vphi)\rightarrow\exp\vphi$, then the universe as a whole evolved into the final stable GR--de Sitter state.

Now we are in position to put together the main results of the present research, in order to get to physical conclusions (we shall include, also, results of \cite{hrycyna, bd-quiros}).

\begin{itemize}

\item It is a very difficult (perhaps hopeless) task to find a BD potential that allows to meet at once the terrestrial/solar system bounds and the bounds coming from cosmological considerations (section \ref{estimate-sec}).  

\item The potential $U\propto\exp\vphi$, is singular in the sense that the dilaton is strictly massless during the course of the entire cosmic history, since the corresponding $U_\text{ch}(\vphi)$ does not develop minimums (see Eq. (\ref{mass-lambda}) with $\lambda=1$). Consequently, the chameleon effect does not arise.

\item Only for the specific exponential potential $U(\vphi)\propto\exp\vphi$, or for potentials that asymptote to $\exp\vphi$, the GR--de Sitter solution is an attractor of the corresponding dynamical system \cite{hrycyna, bd-quiros} (see the section \ref{dyn-syst-sec} of the present paper).

\item In order to get a consistent local picture of transmutation of the Brans--Dicke theory into de Sitter--general relativity (see section \ref{gr-attractor-sec}), assuming that the cosmological dynamics is dictated by the exponential potential $U(\vphi)\propto\exp(\lambda\vphi)$, it is required that $\lambda\gg 1$ (section \ref{ccp-sec}).

\end{itemize}

The conclusion is that, if assume that the BD theory can explain the present acceleration of the expansion within the required accuracy, then the BD scalar field (our dilaton) has had to be detected in solar system and/or in terrestrial experiments. The contrary statement is also true: if the BD theory explains the local gravitational laws and, simultaneously, the dilaton is effectively screened from terrestrial and solar system detection, then it can not explain the present accelerated stage of the cosmic evolution.


\section{acknowledgment}

The authors want to thank SNI-CONACyT for continuous support of their research activity. I Q thanks CONACyT of M\'exico for support of the present research project. The work of R G-S was partially supported by SIP20150188, SIP20144622, COFAA-IPN, and EDI-IPN grants.


\section{Appendix: The thin--shell effect}\label{appendix}

The chameleon effect is primarily linked with the dependence of the effective mass of the BD scalar field upon the density of the surroundings. There is, however, another source of further screening of the fifth-force which can be associated with the BD scalar field. This is called as the thin--shell effect \cite{cham, cham-khoury, cham-tamaki}. This effect, which is significant only for large objects, is originated from the dependence of the strength of the coupling of the BD field to the matter, on the density of the environment as well. As a matter of fact, the BD scalar field outside such a large object, couples to matter with an effective coupling strength $\beta_\text{eff}$ which is much smaller than the ``bare'' coupling $\beta$ (in the case of the BD theory considered in this paper $\beta=1$). In order to consider the classical example \cite{cham-khoury}, imagine an enough large spherical body of mass $M_c$, homogeneous density $\rho_c$ and radius $R_c$, which is immersed in an homogeneous environment of density $\rho_\infty$. Denote by $\phi_c$ and $\phi_\infty$, the values of the BD scalar field which minimize the chameleon potential $V_\text{ch}(\phi)$ for $\rho_c$ and $\rho_\infty$, respectively. The mass of the small perturbations of the BD scalar field about these minima will be denoted by $m_c$ and $m_\infty$, respectively. We have that (see Eq. (\ref{bd-cham-pot})): $$V_\text{ch}(\phi(r))=V_\text{eff}(\phi(r))-\frac{\phi(r)}{3+2\omega_\textsc{bd}}\,\rho(r),$$ where $\rho=\rho_c$ for $r<R_c$, while $\rho=\rho_\infty$ for $r>R_c$. The same for $m_\phi=m_\phi(r)=\sqrt{\der^2_\phi V_\text{ch}(\phi(r))}$: $m_\phi=m_c$ if $r<R_c$, while $m_\phi=m_\infty$ if $r>R_c$. The resulting Helmholtz equation, $$\frac{1}{r^2}\frac{d}{dr}\left[r^2\frac{d(\delta\phi)}{dr}\right]=m^2_\phi\,\delta\phi,$$ for the perturbations around $\phi_c$ and $\phi_\infty$, is subject to the following reasonable boundary conditions: i) the field is regular at the origin, and ii) far away from the body the field tends to its minimum value $\phi_\infty$, $$\left.\frac{d\phi(r)}{dr}\right|_{r=0}=0,\;\lim_{r\rightarrow\infty}\phi(r)=\phi_\infty.$$ The solutions of the Helmholtz equation are \cite{cham-khoury}: $$\phi(r<R_c)=\phi_c+\frac{C_1\sinh(m_c r)}{r},$$ inside of the body, and $$\phi(r\gg R_c)=\phi_\infty+\frac{C_2\,e^{-m_\infty r}}{r},$$ outside it, respectively. Requiring, additionally, continuity of the field and its first derivative at the boundary $r=R_c$, allows to fix the constants $C_1$ and $C_2$. In order to make estimates it is usually assumed that the density contrast between the body and the environment is large: $\phi_c\gg\phi_\infty$, $m_c\gg m_\infty$. After the mentioned assumptions, the exterior solution above can be written as: $$\phi(r>R_c)\approx 2\beta_\text{eff}\frac{M_c\,e^{-m_\infty(r-R_c)}}{r}+\phi_\infty,$$ where $$\beta_\text{eff}=\frac{3\phi_c}{\rho_c R^2_c}=\frac{9}{m_c^2 R_c^2},$$ is the effective coupling of the chameleon field to the surrounding matter. For large objects with $R_c\gg m_c^{-1}$, where $m_c^{-1}$ is the Compton wavelength of the chameleon field, the exterior solution is that of a point particle, but with a much weaker effective coupling to matter $\beta_\text{eff}\ll 1$. This suppression mechanism is what is called as thin--shell effect. The name is motivated by the fact that the BD (chameleon) field $\phi(r)$ is almost a constant throughout the bulk of the object except within a thin shell of thickness $\sim m_c^{-1}$. Only that part of the body located within this thin shell, contributes to the fifth-force exerted on a test particle. This fact results in a very suppressed effective coupling $\beta_\text{eff}\ll 1$.


\end{document}